\documentclass[12pt]{article}
\usepackage[T1]{fontenc}
\usepackage{tikz,pgfplots}
\usetikzlibrary{matrix}
\usepgflibrary{shapes.misc}
\usetikzlibrary{arrows,shadows}
\usepackage{graphicx}
\usepackage[utf8]{inputenc}
\usepackage{amsmath}
\usepackage{mathtools}
\usepackage{amsfonts}
\usepackage{amssymb}
\usepackage{booktabs}
\usepackage[english]{babel}
\usepackage{setspace}
\usepackage{geometry}
\usepackage{xcolor}
\usepackage{color, colortbl}
\usepackage{tikz}
\newcommand{\nn}{\nonumber}
\newcommand{\bea}{\begin{eqnarray}}
\newcommand{\eea}{\end{eqnarray}}    
\newcommand{\be}{\begin{equation}}
\newcommand{\ee}{\end{equation}}    
\newcommand{\UPMNS}{U_{\mathrm{PMNS}}}
\newcommand{\Gr}{\mathcal{G}}
\newcommand{\diag}{\mathrm{diag}}

\newcommand{\thr}{\theta_{13}}
\newcommand{\JCP}{J_{\mathrm{CP}}}
\newcommand{\tha}{\theta_{23}}
\newcommand{\mbf}{\mathbf}
\newcommand{\nline}{ \displaybreak[0]\\}
\newcommand{\ep}{\epsilon}
\newcommand{\mean}[1]{\langle#1\rangle}
\def\dd{\displaystyle}
\def\R{{\rm R}}
\def\L{{\rm L}}
\begin{document}
\thispagestyle{empty}
$\,$

\vspace{32pt}
\begin{center}

\textbf{\Large  GUT and flavor models for neutrino masses and mixing} 

\vspace{30pt}
D. Meloni$^a$
\vspace{16pt}

\textit{$^a$Dipartimento di Matematica e Fisica, 
Universit\`a di Roma Tre\\Via della Vasca Navale 84, 00146 Rome, Italy}\\
\vspace{16pt}

\texttt{davide.meloni@uniroma3.it}
\end{center} 
 \abstract
In the recent years neutrino experiments have studied in detail the phenomenon of neutrino oscillations
and most of the oscillation parameters have been measured with a good accuracy.
However, in spite of many interesting ideas, the
problem of flavor in the lepton sector remains an open issue. In this review, 
we discuss the state of the art of models for neutrino masses and mixing formulated in the context of flavor symmetries, with particular emphasis
on the role played by grand unified gauge groups.

\section{Introduction}
In the course of the last two decades, 
valuable experimental evidences for three families of massive 
neutrinos and flavour neutrino oscillations were obtained in various experimental channels,
and the parameters which characterize the mixing are now known with a relatively high precision. 
As a consequence, the existence of non-vanishing neutrino masses and mixing have been firmly established. 
In spite of the huge amount of available data, many properties of the neutrino physics are yet poorly known or even completely unknown as, just to mention some of them,  
whether the massive neutrinos are Dirac or Majorana particles \cite{Majorana:1937vz}, what kind of spectrum the neutrino masses obeys, what is the absolute scale of neutrino masses, 
what is the octant for the atmospheric mixing angle $\theta_{23}$ and what are the values of the $CP$ violating phases in the leptonic sector.
In a unified description of fermion masses and mixing, the above-mentioned features must be somehow linked to quark properties which, however, appear so dissimilar to make 
such a connection very hard to find; this is the well-known {\it flavor problem}. Let us take the mixing angles as an example. Quark and neutral leptonic mixings are described by the 
Cabibbo-Kobayashi-Maskawa matrix $V_{CKM}$ \cite{Cabibbo:1963yz,Kobayashi:1973fv} and the Pontecorvo-Maki-Nakagawa-Sakata matrix $\UPMNS$ \cite{Pontecorvo:1957cp}-\cite{Pontecorvo:1967fh}, 
respectively.
\begin{figure}[h!]
\begin{center}
\begin{tikzpicture}
\matrix (inputCKM) [matrix of nodes,
                nodes={circle, draw=black, minimum size=.8cm},left delimiter={(},right delimiter={)}] at (0,0)
{
|[fill=black!97]| & |[fill=black!23]|  & |[fill=black!00]|         \\
|[fill=black!23]| & |[fill=black!97]|  & |[fill=black!04]|      \\
|[fill=black!01]| & |[fill=black!04]|  & |[fill=black!99]|         \\
};
\node [draw=none,left=2.7cm, above=1cm]at (inputCKM.south) {$\|V_{CKM}\|=$};

\matrix (inputPMNS) [matrix of nodes,
                nodes={circle, draw=black, minimum size=.8cm},left delimiter={(},right delimiter={)}] at (5.5,0)
{
|[fill=black!79]| & |[fill=black!52]|  & |[fill=black!14]|         \\
|[fill=black!23]| & |[fill=black!45]|  & |[fill=black!62]|      \\
|[fill=black!25]| & |[fill=black!46]|  & |[fill=black!60]|         \\
};
\node [draw=none,left=2.7cm, above=1cm] at (inputPMNS.south) {$\|\UPMNS\|=$};
\end{tikzpicture}
\end{center}

\begin{center}
 \begin{tikzpicture}
 \begin{axis}[
    hide axis,
    scale only axis,
    height=0pt,
    width=0pt,
    colormap={}{ gray(0cm)=(1); gray(1cm)=(0);},
    colorbar horizontal,
    point meta min=0,
    point meta max=1,
    colorbar style={
        width=8cm,
        xtick={0, 0.20, 0.4, 0.6, 0.8, 1}
    }]
\addplot [draw=none] coordinates {(0,0)};
\end{axis}
\end{tikzpicture}
\end{center}

\caption{\it Pictorial representation of the absolute values of the matrix elements of the $V_{CKM}$ and $\UPMNS$ matrices.}
\label{fig:CKM_PMNS}

\end{figure}
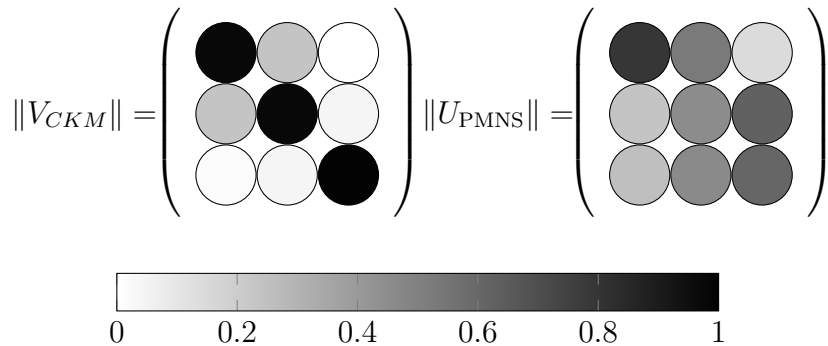
Although one can assume an identical parametrization, Fig.\ref{fig:CKM_PMNS} shows that the absolute values of the matrix elements are quite different: the $V_{CKM}$ is an almost diagonal matrix, with the largest deviation from $1$ coming from the Cabibbo angle in the $(12)$ position  while the $\UPMNS$ exhibits a pattern where all but the $(13)$ entry are of the same order of magnitude of ${\cal O}(1)$. Since at the end of the day the $V_{CKM}$ and $\UPMNS$ matrices all come from the Yukawa matrices of the theory, one would naively expect no sort of relations among their entries, which is obviously the case. Unless one decides to take seriously the numerical {\it quark-lepton complementarity} relation \cite{Raidal:2004iw}-\cite{Frampton:2004vw}
that connects the solar $\theta_{12}$ and atmospheric $\theta_{23}$ leptonic angles to the Cabibbo angle  $\theta_C$, $\theta_{12}+\theta_C\sim \pi/4$. In this case (and also for other similar relations), Grand Unified Theories (GUT) supplemented with the help of family symmetries
could provide a simple explanation so that their role in deciphering the flavor problem cannot be neglected. 
In fact, while GUT groups relate the properties of particles belonging to different species, thus establishing a connections among mass matrices of leptons and quarks, 
flavor symmetries act on the members of particles of the same species but different families, enabling a strong connection between the matrix elements of a given mass matrix. Thus one can arrange the theory in such a way that flavor symmetries are mainly responsible for a definite mixing pattern in the neutrino sector and that GUT symmetries introduce the Cabibbo angle in the leptonic sector as a correction to the $\UPMNS$ given by the diagonalization of the charged lepton mass matrix (somehow related to the down quark masses).

Notice that the additional degree of symmetry involved in these theories allows a substantial decrease of the number of independent parameters compared to the Standard Model case (which amounts to $19$) and, quite often, the model produces observable predictions that can be verified by experiments. The typical example in GUT theories is related to the mean life of the proton $\tau_p$; since the new colored gauge bosons and scalars implied by the larger symmetry can mediate proton decay at a rate faster than the age of the Universe, many variations have been ruled out based on the predicted upper limit on $\tau_p$. 
On the other hand, the less freedom in the elements of the mass matrices subsequent to the imposition of flavor symmetries allowed in the past to derive patterns of leptonic mixing in very good agreement with the old neutrino data which unfortunately do not resist to the comparison with the more precise measurements as we currently have. The typical example is provided by the so-called Tribimaximal mixing (TBM \cite{Harrison:2002er}-\cite{Harrison:2003aw}, more on this and other patterns later in Sect.\ref{sect:flav}) which predicts $\theta_{13}=0$ and requires ad-hoc large corrections to fall over acceptable ranges. 
Given the vastness of the scientific production in terms of models employing flavor symmetries, we restrict ourselves here to non-abelian discrete symmetries and abelian $U(1)$'s. While the latter have been inspired by the Froggatt and Nielsen mechanism \cite{Froggatt:1978nt}, the former answers to the necessity of explaining the existence of three generations of fermions or at least to unify two of them (that is why {\it non-abelian} group), avoiding at the same time the presence of Goldstone and gauge bosons coming from their spontaneous symmetry breaking (that is why {\it discrete}). Discrete symmetries can be inspired by different extensions of the Standard Model (SM); for example, one can start with an $SU(3)$ invariant theory and then break it into its discrete groups using large Higgs representations \cite{Luhn:2011ip}; or one can consider extra dimensional theories \cite{Altarelli:2006kg} (also string inspired), where the new dimensions are properly compactified and the discrete group appears as a remnant of the $n$-dimensional space-time symmetry \cite{Altarelli:2006kg}.

Although the combination $GUT \oplus flavor$ seems to be even more restrictive in terms of free parameters, the aim of this short review is to show that several attempts in this direction have been done that produced good results. But, before arriving at this conclusion, we will devote Sect.\ref{sect:gut} to the understanding of the main prediction for neutrino masses in GUT theories and Sect.\ref{sect:flav} on the role played by  flavor.
Only in Sect.\ref{gutflav} we will investigate the physics opportunity given by the union of these two different types of symmetries.

\section{Remarks on neutrino masses}
\subsection{Dirac mass term}

Dirac neutrino masses can be generated by the same Higgs mechanism that gives masses to quarks and charged leptons in the SM. 
To this aim, we need to introduce SM singlet fermions  $\nu_{Ri}$ and the related Yukawa couplings with the Higgs field; after  spontaneous symmetry breaking, the Lagrangian containing the lepton mass terms is given by:
\begin{equation}
\label{yukawa}
\mathcal{L}_{mass} = - \frac{v}{\sqrt{2}} \sum_{\alpha, \beta = e, \mu, \tau} (\overline{\nu}_{\alpha L} Y^\nu_{\alpha \beta} \nu_{\beta R} + \textrm{h.c.}) - \frac{v}{\sqrt{2}} \sum_{\alpha, \beta = e, \mu, \tau} (\overline{\ell}_{\alpha L} Y^\ell_{\alpha \beta} \ell_{\beta R} + \textrm{h.c.}) \ ,
\end{equation}
where $\ell_{\alpha}$ represents the charged lepton fields, $v$ is the vacuum expectation value (vev) of the Higgs field and $Y^\nu$ and $Y^\ell$ are the Yukawa couplings of neutrinos and charged leptons, respectively, accommodated in  $3 \times 3$ matrices. 
The diagonalization of  $Y^{\nu, \ell}$ can be performed with a biunitary transformation:
\begin{align}
\label{bi_tranf}
{U^\nu_{L}}^\dagger Y^\nu U^\nu_R & = Y^{\prime \nu} \ \quad \textrm{with} \quad Y^{\prime \nu}_{ij} = y^{\prime \nu}_{i} \delta_{ij} \ ,\\ 
{U^\ell_{L}}^\dagger Y^\ell U^\ell_R & = Y^{\prime \ell} \ \quad \textrm{with} \quad Y^{\prime \ell}_{\alpha \beta} = y^{\prime \ell}_{\alpha} \delta_{\alpha \beta} \ ,
\end{align}
and, consequently,  the left  and right-handed components of the fields with definite mass are as follows:
\begin{align}
\label{mass_field}
\nu_{kL} &= \sum_{\beta = e, \mu, \tau} ({U^\nu_L}^\dagger)_{k \beta}\, \nu_{\beta L}  \  , \quad \nu_{kR} = \sum_{\beta = e, \mu, \tau} ({U^\nu_R}^\dagger)_{k \beta} \,\nu_{\beta R} \ , \\
\label{mass_field2}\ell_{\alpha L}^\prime &= \sum_{\beta = e, \mu, \tau} ({U^\ell_L}^\dagger)_{\alpha \beta}\, \ell_{\beta L}  \  , \quad \ell_{\alpha R}^\prime = \sum_{\beta = e, \mu, \tau} ({U^\ell_R}^\dagger)_{\alpha \beta}\, \ell_{\beta R}  \ .
\end{align}
In terms of the mass states defined in eqs.(\ref{mass_field}) and (\ref{mass_field2}), the Lagrangian in  (\ref{yukawa}) can be rewritten as:
\begin{align}
\label{mass_d}
\mathcal{L}_{mass} &= - \sum_{k = 1,2,3} \frac{v y^{\prime \nu}_k}{\sqrt{2}}(\overline{\nu}_{k L} \nu_{k R} + \textrm{h.c.}) - \sum_{\alpha = e, \mu, \tau} \frac{v y^{\prime \ell}_\alpha}{\sqrt{2}}(\overline{\ell}^\prime_{\alpha L} \ell^\prime_{\alpha R} + \textrm{h.c.}) = \\
& = - \sum_{k = 1,2,3} \frac{v y^{\prime \nu}_k}{\sqrt{2}} \overline{\nu}_k \nu_k - \sum_{\alpha = e, \mu, \tau}  \frac{v y^{\prime \ell}_\alpha}{\sqrt{2}} \overline{\ell}^\prime_\alpha \ell^\prime_\alpha \ ,
\end{align}
with
\begin{eqnarray}
\label{filed_mass}\nonumber
\nu_k = \nu_{k L} + \nu_{k R} \ ,\qquad   
\ell^\prime_{\alpha} & = \ell^\prime_{\alpha L} + \ell^\prime_{\alpha R} \ .
\end{eqnarray}
More importantly, the mixings driven by ${U^{\nu,\ell}_L}$ enter in the leptonic charged current expressed in terms of mass eigenstates as
\begin{equation}
\label{curr_cc_mix}
J^{\mu}_{CC} = \sum_{k = 1,2,3} \sum_{\alpha = e, \mu, \tau} \overline{\nu}_{k L} \gamma^\mu ({U^\nu_L}^\dagger U^\ell_L)_{k \alpha} \ell^\prime_{\alpha L} \ ,
\end{equation}
and give rise to the  well known PMNS matrix: 
\begin{equation}
\label{eq:mix}
\UPMNS = {U^\ell_L}^\dagger U^\nu_L \ .
\end{equation}
This unitary matrix is generally parametrized in terms of three mixing angles and one CP-violating  phase, in a way 
similar to that used for  $V_{CKM}$:
\begin{align}
\label{PDG}
\UPMNS = \left(\begin{array}{ccc} c_{12}c_{13} & s_{12}c_{13} & s_{13}e^{-i\delta} \\ -s_{12}c_{23}-c_{12}s_{23}s_{13}e^{i\delta} & c_{12}c_{23}-s_{12}s_{23}s_{13}e^{i\delta} & s_{23}c_{13}\\ s_{12}s_{23}-c_{12}c_{23}s_{13}e^{i\delta} & -c_{12}s_{23}-s_{12}c_{23}s_{13}e^{i\delta} & c_{23}c_{13} \end{array} \right )\,,
\end{align}
where $c_{ij} = \cos(\theta_{ij})$, $s_{ij} = \sin(\theta_{ij})$ and $\theta_{ij}$ are the mixing angles ($0\leq\theta_{ij}\leq\pi / 2$). $\delta$ is the Dirac $CP$-violating phase ranging in the interval $0 \leq \delta < 2 \pi$. 
%
%

The current best-fit values and the allowed  1$\sigma$ and 3$\sigma$ ranges for the oscillation parameters as well as for the two independent mass differences $\Delta m_{kj}^2=m_k^2 - m_j^2$, as obtained  from the flavor transition experiments, are summarized in Tab.\ref{tab:oscillation_exp}. Normal Ordering refers to the situation in which $m_1 < m_2 <m_3$, whereas for the Inverted Ordering we mean $m_3 < m_1 < m_2$.

The reported values are obtained from the global analysis of Ref. \cite{Esteban:2016qun}. 
\begin{table}[h!]
\begin{center}
\begin{tabular}{c  c  c  c c}
\toprule
\toprule
& \multicolumn{2}{c}{\bf Normal Ordering} & \multicolumn{2}{c}{\bf Inverted Ordering}\\
\midrule
\bf Parameter &  \bf Best Fit & \bf 3$\sigma$ Range &  \bf Best Fit & \bf 3$\sigma$ Range\\ 
\midrule
$\sin^2\theta_{12}/10^{-1}$ & 3.06$^{+0.12}_{-0.12}$ 	 & 2.71 $\div$ 3.45 	& 3.06$^{+0.12}_{-0.12}$ 	& 2.71 $\div$ 3.45 \\
$\sin^2\theta_{13}/10^{-2}$ & 2.166$^{+0.075}_{-0.075}$ & 1.934 $\div$ 2.392 	& 2.179$^{+0.076}_{-0.076}$ 	& 1.953 $\div$ 2.408 \\
$\sin^2\theta_{23}/10^{-1}$ & 4.41$^{+0.27}_{-0.21}$ 	 & 3.85 $\div$ 6.35 	& 5.87$^{+0.20}_{-0.24}$ 	& 3.93 $\div$ 6.40\\
 $\delta$ 	     & 4.56$^{+0.89}_{-1.03}$  	 & 0 $\div$ 2$\pi$ 	& 4.83$^{+0.70}_{-0.80}$  	& 0 $\div$ 2$\pi$ \\
$\Delta m^2_{21}/10^{-5} \ [\textrm{eV}^2]$ & 7.50$^{+0.19}_{-0.17}$ & 7.03 $\div$ 8.09 & 7.50$^{+0.19}_{-0.17}$ & 7.03 $\div$ 8.09\\
$\Delta m^2_{3\ell}/10^{-3} \ [\textrm{eV}^2]$ & +2.524$^{+0.039}_{-0.040}$ & +2.407 $\div$ +2.643 & -2.514$^{+0.038}_{-0.041}$ & -2.635 $\div$ -2.399 \\
\bottomrule
\bottomrule
\end{tabular}
\caption{\it Value of the oscillation parameters obtained from a global analysis from Ref. \cite{Esteban:2016qun}. For the squared mass difference in the last line, $\ell = 1$ in the Normal Ordering and $\ell = 2$ in the Inverted Ordering.}
\label{tab:oscillation_exp}
\end{center}
\end{table}

\subsection{Majorana mass terms}
\begin{figure}[h!]
\centering
\includegraphics[scale=0.8]{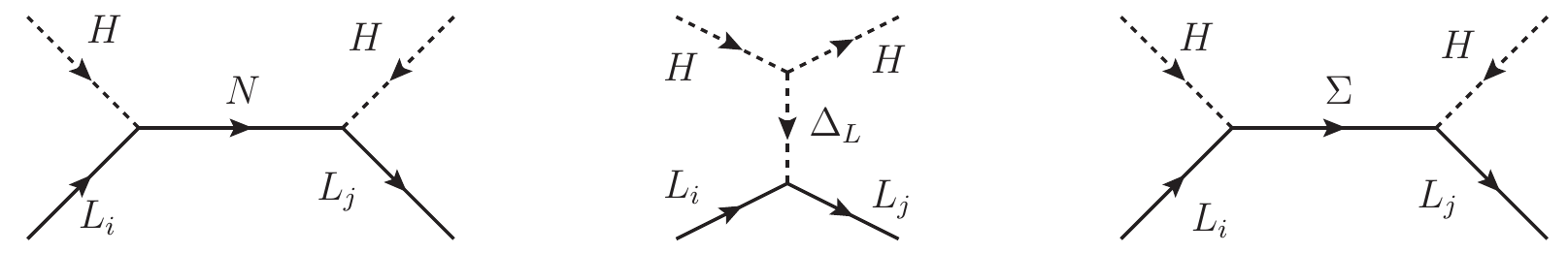}
\caption{\it Tree level realization of the Weinberg operators $O_1$, $O_3$ and $O_4$. From left to right, the intermediate states are: singlet fermion $N$, scalar triplet $\Delta_L$ and 
fermion triplet $\Sigma$ fields. }
\label{RWZ}
\end{figure}

With the minimal particle content of the SM, namely leptons $L_i$ and the Higgs doublet $H$:
\begin{equation}
\label{matrmassa}
L_i=\begin{pmatrix}
\nu \\ 
e
\end{pmatrix}_{iL}\,, \qquad H =
\begin{pmatrix}
\phi^+ \\ 
\phi^0
\end{pmatrix}\,,
\end{equation}
one can generate dimension five operators of the form:
\begin{equation}
\label{weinberg}
 {\cal L}_5 \sim \frac{y_{ij}}{\Lambda}\, \overline L_i L_j^c \tilde H \tilde H^T\,,
\end{equation}
where $\Lambda$ can be understood as the scale where new physics probably sets in and $\tilde H = -i\,\tau_2 H^*$. In fact, two
SM singlets are built from the product of four $SU(2)_L$ doublets as \cite{Ibarra:2016qnf}: 
\begin{eqnarray}
2 \otimes 2 \otimes 2 \otimes 2 = (3 \oplus 1) \otimes (3 \oplus 1)\,,
\end{eqnarray}
either via the product of two triplets or by the product of two singlets. 
Since $L$ and $H$ are different fields, we have four possible combinations that can give an overall $SU(2)_L$ singlet:
\begin{eqnarray}
 O_1 = \left(L_i H\right)_1 \left(L_j H\right)_1 \qquad  & &  O_2 = \left(L_i L_j\right)_1 \left(H H\right)_1 \nonumber \\ 
 O_3 = \left(L_i L_j\right)_3 \left(H H\right)_3  \qquad & &  O_4= \left(L_i H\right)_3 \left(L_j H\right)_3  \,, \nonumber
\end{eqnarray}
where the subscript $1,3$ refer to the $SU(2)_L$ representation. Since $\left(H H\right)_1 = 0 $ due to the antisymmetry
under the exchange of the two doublets, only $O_{1,3,4}$ contribute to neutrino masses. In particular,  the explicit form of the bilinear 
are as follows:
\begin{eqnarray}
\left(L_i L_j\right)_1  \sim \nu_i e_j -   e_i \nu_j     
&\qquad&   \left(L_i L_j\right)_3 \sim     
\begin{pmatrix}
\nu_i \nu_j \\ 
\nu_i e_j + e_i \nu_j\\
e_i e_j
\end{pmatrix}    \\                                                                                                                                                                                                                                                                   
\left(L_i H\right)_1    \sim \nu_i \phi^0 -   e_i \phi^+     
&\qquad&   \left(L_i H\right)_3 \sim     
\begin{pmatrix}
\nu_i \phi^+  \\ 
\nu_i \phi^0 + e_i \phi^+ \\
e_i \phi^0
\end{pmatrix}    \\ &&
\left(H H\right)_3    \sim     
\begin{pmatrix}
\phi^+ \phi^+  \\ 
\phi^+ \phi^0 + \phi^0 \phi^+ \\
\phi^0 \phi^0
\end{pmatrix}   \,,
\end{eqnarray}
from which we realize that $O_1$, $O_3$ and $O_4$ all contain the combination of fields $\nu_i \nu_j (\phi^{0})^2$ that generate neutrino masses after electroweak spontaneous  symmetry breaking.
However, giving their different contractions of the $SU(2)_L$ indices, $O_1$ has a tree-level realization in terms of the interchange of a heavy SM singlet $\nu_R$, the type-I see-saw mechanism  
\cite{Minkowski:1977sc}-\cite{Mohapatra:1980}, whereas heavy triplets are needed to realize $O_3$ and $O_4$, either with the interchange of a scalar particle (the type-II see-saw mechanism \cite{Konetschny:1977bn}) or 
of a fermion field (the type-III  mechanism \cite{Foot:1988aq}), see Fig.(\ref{RWZ}).

In the first case, the introduction of three right-handed neutrinos $N_i \equiv \nu_{R_i}$ allows for an  
invariant mass Lagrangian of the form \cite{Altarelli:2004za}:
\begin{equation}
 {\cal L}_m=- Y_{ij}\bar L_i (\tilde{H} N_{j})+\frac{1}{2}
{\bar N^c}_{i} M_{ij} N_{j} +~h.c.\,.
\label{lag}
\end{equation} 
The first term in this equation is known as the {\it Dirac mass term} and it is essentially a copy of the mass term "employed" by the charged fermions and quarks to get their masses.
The second term, instead, is a pure {\it Majorana} contribution to the neutrino mass.
After spontaneous symmetry breaking, ${\cal L}_m$ gives rise to
the Dirac mass matrix $(m_D)_{ij}\equiv Y_{ij} \langle H \rangle$, which is 
non-hermitian and non-symmetric, and to the Majorana mass matrix $M$ which is symmetric. 
Assuming all $N_i$ to be very heavy, one can integrate them away so that  
the resulting light neutrino mass matrix reads:
\begin{eqnarray}  
\label{seesawI}
m_{\nu}=- m_D M^{-1}m_D^T\,.
\end{eqnarray}
The type-I see-saw mechanism shows that the light neutrino masses depend quadratically on the Dirac
masses but are inversely proportional to the large Majorana mass, so that the scale of new physics is clearly $\Lambda = M$.

In the case of type-II mechanism, at least one scalar $SU(2)_L$ triplet must be added to the field content of the SM; 
for values of the weak hypercharge equal to $+1$, the triplet has the following components:
\begin{eqnarray}
\Delta_L =    
\begin{pmatrix}
\Delta^{++} \\ 
\Delta^{+}\\
\Delta^0
\end{pmatrix}  \,,
\end{eqnarray}
and the Lagrangian terms that accommodate the new states and are relevant for neutrino masses are:
%
\begin{eqnarray}
{\cal L}_{\Delta} \sim \left(k_{ij} \bar L_i \left(\sigma \cdot \Delta_L^\dagger\right) L^c_j  - \mu_\Delta \tilde H^T \left(\sigma \cdot \Delta_L\right) \tilde H + h.c.\right) + m^2_\Delta |\Delta_L|^2\,,
\end{eqnarray}
where $\sigma_i$ are the Pauli matrices and $k_{ij}$ the new Yukawa couplings induced by the presence of $\Delta_L$.
Assuming that the scalar potential has a minimum in the direction $\langle \Delta_L \rangle=(0,0,v_\Delta)$ (as well as in the standard vacuum
$\langle H \rangle=(0,v)$) and that the hierarchy $m^2_\Delta \gg \mu_\Delta v$ is valid, then the light neutrino mass matrix is:
\begin{eqnarray}
 (m_\nu)_{ij} \sim \frac{\mu_\Delta v^2}{m_\Delta^2} k_{ij}\,;
\end{eqnarray}
in this case, the scale of new physics is approximately given by $\Lambda \sim m_\Delta^2/\mu_\Delta$.

In the last case of type-III see-saw mechanism, the triplet hyperchargeless fermions $\Sigma$ can be arranged in the following form:
\begin{eqnarray}
\Sigma =    
\begin{pmatrix}
\Sigma^{0}/\sqrt{2} & \Sigma^{+} \\ 
\Sigma^{-} & -\Sigma^{0}/\sqrt{2}
\end{pmatrix}\,,
\end{eqnarray}
and the related Lagrangian reads:
\begin{eqnarray}
{\cal L}_{\Sigma} -\sim k^\Sigma_{ij}\, \bar L_i \tilde H \Sigma_j   + (m_\Sigma)_{ij} Tr\left(\bar \Sigma^c_i \Sigma_j \right)\,,
\end{eqnarray}
where again  $ k^\Sigma_{ij}$ is a  Yukawa coupling matrix. Under the hypothesis that $m_\Sigma \gg k^\Sigma v$, the light mass matrix assumes the form
\begin{eqnarray}
 m_\nu \sim -k^\Sigma \frac{1}{m_\Sigma} \left(k^\Sigma\right)^T v^2\,,
\end{eqnarray}
which is very similar to eq.(\ref{seesawI}) since, for the purposes of neutrino masses, the state $\Sigma^0$ acts like a right-handed neutrino.

It has to be noted that the Majorana nature of neutrinos modifies the  PMNS matrix of eq.(\ref{PDG}) to take into account two more independent CP violating phases $\alpha$
and $\beta$ that cannot be eliminated by a rotation of the neutrino fields; a possible  convention for the {\it new} $\UPMNS$ is as follows:
\begin{align}
\label{PMNS_definition_general}
 \UPMNS^\prime  = \UPMNS \times \diag\{1, e^{i \alpha/2}, e^{i \beta/2}\}\,.
\end{align}
Neutrino oscillation data cannot determine whether the massive neutrinos are Dirac or Majorana particles because the new phases cancel out of the oscillation amplitudes.

\section{Neutrino masses and mixing in GUT theories}
\label{sect:gut}
The possibility to generate non–zero neutrino masses through the see–saw
mechanism, which requires quite a large $B-L$ scale, fit rather naturally in grand unified models based on the gauge group SO(10) \cite{Fritzsch:1974nn}. 
Putting aside Supersymmetry (SUSY) for the moment,
the experimental constraints from the lifetime of the proton and from the weak mixing angle
$\sin^2 \theta_W$ impose  that SO(10) breaks to the SM
at least in two or more steps \cite{delAguila:1980qag}-\cite{Deshpande:1992au}. 
In a minimal setup which allows for a two-step breaking, the intermediate gauge groups (typically a Pati-Salam group $SU(4) \times SU(2)_L\times SU(2)_R \equiv 4_{C}\, 2_{L}\, 2_{R}$ \cite{Pati:1974yy})
is broken down to the SM  at a  scale around $10^{12}$ GeV, which is usually also the scale of the Majorana masses. 
To accomplish this program, the Higgs sector must be carefully chosen in such a way to avoid bad mass relations of the $SU(5)$ type \cite{Georgi:1974sy}.
Let us discuss an example. Consider the following chain: 
\begin{eqnarray}  
\label{chain}
SO(10)&\stackrel{M_U-{\bf 210}_H}{\longrightarrow}
&4_{C}\, 2_{L}\, 2_{R}\ \stackrel{M_I-{\bf 126}_H}{\longrightarrow}SM\ \stackrel{M_Z-{\bf 10}_H}{\longrightarrow} \ SU(3)_{C}\,U(1)_{EM}
\,,
\end{eqnarray}
where the three mass scales refer to the scale where $SO(10)$ is broken down to the PS ($M_U$), where PS is broken to the SM ($M_I$) and finally where the SM group is broken down to the electromagnetism ($M_Z$). The $SO(10)$ representations used to perform the various stages of symmetry breaking are also indicated.
With fermions in the ${\bf 16}$ representation, the Yukawa Lagrangian contains two terms:
\bea
{\cal L} = {\bf 16} \left(h\, {\bf 10}_H + f\, {\bf \overline{126}}_H\right){\bf 16} \,,
\eea
where the couplings $h$ and $f$ are $3 \times 3$ symmetric matrices in flavor space. 
In terms of their PS quantum numbers, the Higgses in eq.(\ref{chain}) decompose as:
\begin{align}
& {\bf 10}_H = (1,2,2) \oplus (6,1,1) \, ,\nn \\
& {\bf 126}_H = (6,1,1) \oplus (\overline{10},3,1) \oplus (10,1,3) \oplus (15,2,2) \, .\nn
\end{align}
Of all the previous sub-multiplets, the ones useful for generating neutrino (and fermion) masses are
the $(1,2,2) \equiv \Phi \in {\bf 10}_H$ entering the last breaking in eq.(\ref{chain}) and that contains an $SU(2)_L$ doublet, 
the $(10,1,3)\equiv \Delta_R \in {\bf 126}_H$ to allow for right-handed Majorana masses and the 
$(15,2,2)\equiv \Sigma \in  {\bf 126}_H$ which also contains an  $SU(2)_L$ doublet. Using the extended survival hypothesis \cite{delAguila:1980qag}, we assume that both $\Delta_R$ and $\Sigma$  have masses around $M_I$, and all other multiplets 
are close to the GUT scale \footnote{One can safely estimate that the colored states $\Delta_R$ and $\Sigma$ do not give a catastrophic contribution to proton decay \cite{Altarelli:2013aqa,Deppisch:2017xhv}.}.

A comment here is in order. The $(1,2,2)$ of the ${\bf 10}_H$ representation can be decomposed 
into 
\begin{equation}
(1,2,2) = (1,2,+\tfrac{1}{2})\oplus (1,2,-\tfrac{1}{2}) \equiv H_u  \oplus H_d
\end{equation}
under the SM group; if
$10_H = 10_H^\ast$ then $H_u^\ast = H_d$ as in the SM but, as it has been shown in \cite{Bajc:2005zf}, in the limit 
$V_{cb}=0$ the ratio $m_t/m_b$ should be close to $1$, in contrast with the experimental fact 
that at the GUT scale $m_t/m_b\gg 1$. On the other hand, even though the ${\bf 10}_H$ is a real representation from the $SO(10)$ point of view,
one can choose its components to be either real or complex. In the latter case, $10_H \neq 10_H^\ast$ and then $H_u^\ast \neq H_d$. In order to keep the parameter space at an acceptable level, it is a common practice to introduce an extra symmetry (for instance,  the Peccei-Quinn $U(1)_{PQ}$ \cite{Peccei:1977hh}) to avoid the Yukawa couplings related to
$10_H^\ast$. 

For the vev values of the ${\bf 10}_H$ components we will use the following short-hand notation:
\bea
k_u\equiv\langle(1,2,2)_{10}^u\rangle \ne k_d\equiv\langle(1,2,2)_{10}^d\rangle\,.
\label{vev1}
\eea
For the vev of the ${\bf 126}_H$, instead, one can take full advantage of the fact that a vev  for the 
doublet  $\Sigma$ (that we call $v_{u,d}$) can be induced by a term in the scalar potential of the form \cite{Babu:1992ia}:
\bea
\nn V = \lambda\,{\bf 126}_H\,\overline{{\bf 126}_H}\,{\bf 126}_H\,{\bf 10}_H \to \lambda\,\Delta_R\,\overline{\Delta_R}\,\Sigma \,\Phi\,,
\eea
which gives: 
\bea
v_{u,d} \sim \lambda \frac{v_R^2}{M^2_{(15,2,2)}}\,k_{u,d}\,,
\eea
where $v_R=\langle (10,1,3) \rangle$. According to this, the fermion mass matrices of the model assume the form:
\bea
M_u = h\,k_u + f\,v_u, &\qquad&  M_d = h\,k_d + f\,v_d \nn \\
&  \label{masses} & \\
M^D_\nu = h\,k_u -3\, f\,v_u , &\qquad& M_l = h\,k_d -3\, f\,v_d, \qquad M^M_\nu = f\,v_R\,.\nn
\eea

These relations clearly show why the Yukawa sector requires more than the ${\bf 10}_H$; in fact, in the absence of the  ${\bf 126}_H$ (or ${\bf 120}_H$) one would get 
$M_d \equiv M_l$, which is phenomenologically wrong. The role of the ${\bf 126}_H$ in $SO(10)$ theories is exactly to break the wrong mass relations and the factor of $3$ appearing in  eq.(\ref{masses}), derived from the vev of $\Sigma$ of the  ${\bf 126}_H$, is the equivalent of the Georgi-Jarlskog factor of the non-minimal $SU(5)$ \cite{Georgi:1979df}.

Under the hypothesis that the type-I see-saw mechanism is responsible for the light neutrino masses, a fit can be performed which fixes the entries of the $h$ and $f$ couplings to reproduce the low energy observables in the flavor sector (also in the supersymmetric case) in the full three-flavor approach \cite{Altarelli:2013aqa,Joshipura:2011nn,Dueck:2013gca}. This partially contradicts the conclusions derived in the two-flavor limit, where the type-I see-saw mechanism  has been shown to be incompatible with a large atmospheric mixing. To show this, let us approximate $M_\nu^D \approx M_u$ and work in the basis where the charged leptons are diagonal; assuming a small up and down quark mixings $\lambda_C$ (of the order of the Cabibbo angle), eq.(\ref{seesawI}) tells us that
\begin{eqnarray}
m_\nu \sim 4\, r_R\,   
\begin{pmatrix}
m_c^2/(m_s - m_\mu) & \lambda_C \\ 
\lambda_C &  m_t^2/(m_b - m_\tau)
\end{pmatrix}\,, 
\end{eqnarray}
so that two non-degenerate eigenvalues can be generated whose squared difference can be made of the correct order of magnitude $\sim 10^{-3}$ $eV^2$, but the atmospheric mixing angle is suppressed by $\lambda_C$, thus making this construction incompatible with the data.

Relations of the form (\ref{masses}) are also obtained in the minimal $SU(5)$ scenario with a ${\bf 5}_H$ and fermions in the reducible $\bar {\bf 5} \oplus {\bf 10}$ representation. 
With this minimal Higgs content, the prediction at the GUT scale is again  $M_d \equiv M_l$. To solve this problem, the scheme proposed in \cite{Georgi:1979df} involved a slightly more complicated Higgs structure due to the presence of the ${\bf 45}_H$ representation. It replaces the above wrong relations with the more appropriate $m_d = 3 m_e$ and $3 m_s = m_\mu$, which can be derived from the following textures \cite{Arason:1992eb}:
\begin{eqnarray}
\label{textures}
Y_u =    
\begin{pmatrix}
0 & p &0 \\ 
p & 0 & q \\
0 & q & v
\end{pmatrix}\,, \quad 
Y_d =    
\begin{pmatrix}
0 & r &0 \\ 
r & s & 0 \\
0 & 0 & t
\end{pmatrix}\,, \quad 
Y_e =    
\begin{pmatrix}
0 & r &0 \\ 
r & -3s & 0 \\
0 & 0 & t
\end{pmatrix}\,,
\end{eqnarray}
and whose flavor structure can be obtained, for example, by means of additional symmetries (discussed later). 
In the context of $SO(10)$, the textures in eq.(\ref{textures}) have been obtained in \cite{Harvey:1980je,Harvey:1981hk}, in a model with three 
families of left-handed fermions, $16_{1,2,3}$,  
two real ${\bf 10}_H$'s, three ${\bf 126}_H$ and one ${\bf 45}_H$. 
Equally successful phenomenological attempts where instead all quark and lepton mass matrices have the
same zero texture with vanishing (1,1), (1,3) and (3,3) entries have been proposed in \cite{Matsuda:1999yx}.

Going beyond the type-I see-saw mechanism for neutrino masses, 
it has been shown that there exists a very elegant connection between the large atmospheric angle $\theta_{23}$ and the relation $m_b = m_\tau$, if the type-II see-saw is the dominant one
\cite{Bajc:2001fe,Bajc:2002iw}. To show this, let us allow the 
$(\overline{10},3,1)$ component of the ${\bf 126}_H$ to take a large vev $v_L$. This generates a "left" mass matrix for the Majorana neutrinos $M^L_\nu = f\,v_L$ so that the total light neutrino mass matrix is given by $m_\nu =M^L_\nu - m_D^T (M_\nu^M)^{-1}m_D $. Under the hypothesis of the dominance of type-II, in the basis where the charged leptons are diagonal we easily get:
\bea
m_\nu = M^L_\nu \approx M_d - M_l \approx 
\begin{pmatrix}
m_s - m_\mu & \theta_D \\ 
\theta_D &  m_b - m_\tau
\end{pmatrix}\,,
\eea
($\theta_D$ being a small down quark mixing)
and a maximal atmospheric mixing necessarily requires a cancellation between $m_b$ and $m_\tau$. However, SM extrapolation of the fermion masses from the electroweak scale up to the GUT scale (but see \cite{Meloni:2014rga,Meloni:2016rnt} for the effects of the intermediate mass scales in the running) shows that $m_b \sim 1.7 m_\tau$  \cite{Xing:2007fb},  so this mechanism does not seem to fit well with a non-SUSY $SO(10)$ GUT with the ${\bf 10}_H\oplus {\bf 126}_H$ Higgs sector \cite{Bertolini:2006pe}. This conclusion is not altered  when the fit takes into account the three families of fermions.
On the other hand, in the SUSY case the relation $m_b=m_\tau$ is roughly fulfilled for low $\tan \beta\sim{\cal O}(1)$ with no threshold corrections but also for larger $\tan \beta\sim{\cal O}(40)$ with significant threshold corrections. The quality of the full three-family fits in these cases is comparable.

If we insist on minimality in the Higgs sector, the next combinations are the   ${\bf 120}_H\oplus {\bf 126}_H$ and ${\bf 10}_H\oplus {\bf 120}_H$. Both of them make use of the ${\bf 120}_H$ representation which, according to the following decomposition under the PS gauge group, contains several bi-doublets useful for fermion masses:
\begin{align}
& {\bf 120}_H = (10+\overline{10},1,1)\oplus (6,3,1)\oplus (6,1,3)\oplus (15,2,2) \oplus (1,2,2) \, .\nn 
\end{align}
Models of the first kind  (${\bf 120}_H\oplus {\bf 126}_H$) have been considered predictive
when restricted to the second and third generations \cite{Bajc:2005zf}. However, the predicted ratio $m_b/m_\tau\sim 3$ strongly disfavors a SM (for which $m_b/m_\tau\sim 2$) and SUSY (for which $m_b/m_\tau\sim 1$) fits with neither type-I nor type-II see-saw dominance.
The second combination, ${\bf 10}_H\oplus {\bf 120}_H$ \cite{Lavoura:2006dv}, in spite of being compatible with the $b-\tau$ unification \cite{Bajc:2005aq}, produces either  down-quark mass or  top-quark mass unrealistically  small.

In the case of a non-minimal Higgs content with  ${\bf 10}_H\oplus {\bf 120}_H\oplus {\bf 126}_H$, the Yukawa sector contains a large number of independent parameters but,  
except the supersymmetric case, the use of the ${\bf 120}_H$  does not  improve the fits in the type-II see-saw dominated case. On the other hand, 
the fits obtained for the type-I scenario, including neutrino observables, are
considerably better than the corresponding SUSY  as well as better of the ${\bf 10}_H\oplus {\bf 126}_H$
non-SUSY case.

\section{Neutrino masses and mixing from flavour symmetries}
\label{sect:flav}
\subsection{Lepton mixing from discrete symmetry}
\label{sec:lepton_mixing}
\begin{figure}[h]
	\begin{center}
		\begin{tikzpicture}[sibling distance=10em,
		every node/.style = {shape=rectangle, rounded corners,
			draw, align=center,
			top color=white, bottom color=black!10}]]
		\node (top) {Flavour symmetry \textcolor{blue}{$\Gr_f$}}  
		child { node (Gl) {Charged Lepton\\Sector \textcolor{blue}{$\Gr_\ell\subset \Gr_f$}}
			child { node (Q) {$Q^\dagger M_\ell^\dagger M_\ell Q = M_\ell^\dagger M_\ell$\\$Q \in \Gr_\ell$} 
				child { node (Qdiag) {$Q^{\diag} = U_\ell^\dagger Q U_\ell$} }}}
		child { node (Gnu) {Neutrino Sector\\ \textcolor{blue}{$\Gr_\nu\subset \Gr_f$}}
			child { node (Z) {$Z^T M_\nu Z = M_\nu$\\$Z \in \Gr_\nu$} 
				child { node (Zdiag) {$Z^{\diag} = U_\nu^\dagger Z U_\nu$} }}};
		\path (Zdiag) -- (Qdiag) node[pos=.5,below=1cm] (PMNS) {$\textcolor{red}{\UPMNS = U_\ell^\dagger U_\nu}$};
		\draw[-latex,bend left=30]  (Zdiag) edge (PMNS);
		\draw[-latex,bend right=30]  (Qdiag) edge (PMNS);
		\end{tikzpicture}
		\caption{\it Representative scheme of the approach used to construct the $\UPMNS$.} \label{fig:scheme}
	\end{center}
\end{figure}
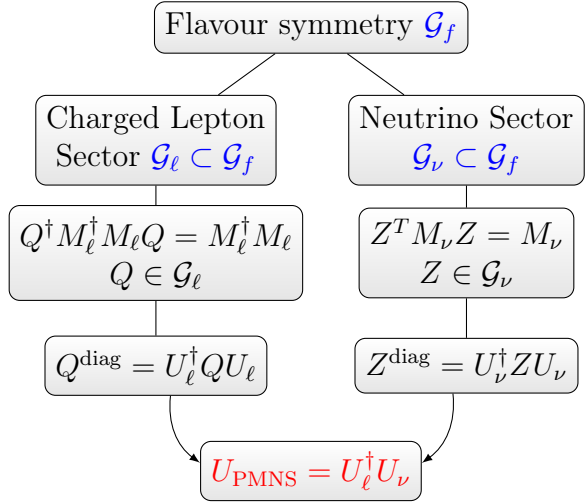

The general strategy to get the leptonic mixing matrix $\UPMNS$ from symmetry consideration is to assume that at some large energy scale the theory is invariant under the action of a flavour symmetry group $\Gr_f$; the scalar sector is then built in a suitable way as to be broken to different subgroups in the neutrino sector $\Gr_\nu$, and in the charged lepton sector, $\Gr_\ell$. The lepton mixing originates then from the mismatch of the embedding of $\Gr_\ell$ and $\Gr_\nu$ into $\Gr_f$. 
Let us assume that
\begin{align}
\Gr_\ell \subset \Gr_f \qquad \Gr_\nu \subset \Gr_f \qquad \Gr_\ell \cap \Gr_\nu = \emptyset .
\end{align}
For Majorana particles, we can write the action of the elements of the subgroups of $\Gr_f$ on the mass matrix as \footnote{The charged lepton mass matrix $M_\ell$ is written in the right-left basis.}
\begin{subequations}
	\label{mass_invariance_conditions}
	\begin{gather}
	Q^\dagger M_\ell^\dagger M_\ell Q = M_\ell^\dagger M_\ell\qquad Q \in \Gr_\ell\\
	Z^T M_\nu Z = M_\nu \qquad Z \in \Gr_\nu\,.
	\end{gather}
\end{subequations}
For Dirac neutrinos the last relation must be modified as:
\begin{align}
Z^\dagger M_\nu^\dagger M_\nu Z = M_\nu^\dagger M_\nu \qquad Z \in \Gr_\nu.
\end{align}
If we restrict ourselves to matrices $Z$ with $\det Z=1$ and to Majorana neutrinos, then 
the maximal invariance group of the neutrino mass matrix which leave the neutrino masses unconstrained is the Klein group $V = Z_2 \otimes Z_2$ \cite{Lam:2007qc,Lam:2008rs,Lam:2008sh,Fonseca:2014koa}. 
The charged leptonic subgroup $\Gr_\ell$ could be either a cyclic group $Z_n$, with the index $n\ge 3$, or a product
of cyclic symmetries like, for example, $Z_{2} \otimes Z_{2}$. We discard in the discussion possible residual non-abelian symmetries because their character would result in a partial or complete  degeneracy of the mass spectrum, and thus incompatible with the current data on charged lepton masses. For the same reason we assume that $Z \in \Gr_\nu$ decomposes into three inequivalent representations under $\Gr_\ell$.\\  
The diagonalization of the mass matrices is equivalent, using \eqref{mass_invariance_conditions}, to a  rotation of the group elements $Q$ and $Z$ through unitary matrices as:
\begin{subequations}
	\begin{align}
	Q^{\diag} &= U_\ell^\dagger Q U_\ell \\
	Z^{\diag} &= U_\nu^\dagger Z U_\nu\,,
	\end{align}
\end{subequations}
because both $\Gr_\ell$ and $\Gr_\nu$ are abelian. 
The matrices $U_\ell$ and $U_\nu$ are determined up to unitary diagonal $K_{\ell, \nu}$ and permutation $P_{\ell, \nu}$ matrices:
\begin{subequations}
	\begin{align}
	U_\ell &\longrightarrow U_\ell P_\ell K_\ell \\
	U_\nu &\longrightarrow U_\nu P_\nu K_\nu.
	\end{align}
\end{subequations}
Thus, up to Majorana phases and permutations of rows and columns, the lepton mixing matrix $\UPMNS$ is given by:
\begin{align}
\label{PMNSdefinition}
\UPMNS = U_\ell^\dagger U_\nu.
\end{align}
Notice that, as a consequence of the fact that  $\UPMNS$ is not completely determined, the mixing angles are fixed up to a small number of degeneracies. For the same reason, the Dirac $CP$ phase $\delta$ is determined up to a factor $\pi$ and the Majorana phases cannot be predicted  because the matrix $M_\nu$ remains unconstrained in this setup. 
In Fig. \ref{fig:scheme} we have pictorially summarized the above procedure.

It is remarkable that, under particular assumptions on the residual symmetry groups in the neutrino and charged lepton sectors\footnote{For instance, one can impose relations between the generators of these residual groups and/or force the determinants to assume specific values.}, 
the construction we have just discussed allow for model (and mass)-independent predictions on the mixing angles (or columns of $\UPMNS$).
As it has been shown in \cite{Hernandez:2012ra,Hernandez:2012sk,Grimus:2013rw}, if only a cyclic group from each sector is a subgroup
of the full flavor group $\Gr_f$, then it is possible to derive non-trivial relations between the mixing matrix in
terms of the symmetry transformations which, in turn, provoke  the appearance of well-defined connections among different mixing angles, also called {\it sum rules}.
In particular, non-zero $\theta_{13}$, deviations from maximal mixing for $\theta_{23}$  and predictions for the CP Dirac phase \cite{Ge:2011ih,Ge:2011qn}
are relevant predictions in (quasi perfect) agreement with the current data. An intersting and useful  classification of all possible mixing matrices completely determined by residual
symmetries (originated from a finite flavour symmetry group) can be found in \cite{Fonseca:2014koa}.

Since the family symmetry $\Gr_f$ has to be broken to generate the observed pattern of masses and mixing, the models generally consider an enlarged Higgs sector where 
Higgs-type fields, called {\it flavons} $\phi$, are 
neutral under the SM gauge group and break spontaneously the family symmetry by acquiring a vev
\begin{equation}
 \epsilon= \frac{\langle \phi \rangle }{\Lambda}\ ,
\end{equation}
where $\Lambda$ denotes a high energy mass scale. If the scale of the vev is smaller (or at least of the same order of magnitude) than $\Lambda$, one can consider $\epsilon$ as a small expansion parameter which can be used to derive Yukawa matrices with built-in hierarchies and/or precise relations among their entries. In order to do that, it is often necessary that 
all three lepton families are grouped into triplet irreducible representations, so that the possible choices for $\Gr_f$ are $U(3)$ and subgroups. 
To give an example,  in the case of $SU(3)$ and for the Weinberg
operator of eq.~(\ref{weinberg}), one can consider lepton doublets into a triplet of $SU(3)$ and the Higgs doublet $H$ 
in a singlet of $\Gr_f$ \cite{King:2001uz,King:2013eh};
the lowest dimensional $SU(3)$ invariant
operator is built using a pair of flavon fields transforming in the ${\bf \overline{3}}$ of $SU(3)$.
For a generic flavon alignment  $\langle \phi \rangle \propto (a,b,c)^T$,  the neutrino mass matrix is then proportional to 
\begin{equation}
\begin{pmatrix}
a^2&ab&ac \\
ba & b^2 & bc \\
ca&cb&c^2
\end{pmatrix} \ .
\end{equation}
Special mixing patterns, as the ones discussed below, are obtained assuming particular flavon alignments in the flavor space which, quite frequently, imply well defined relations among the mixing angles and the Dirac CP-violating phase \cite{Barry:2010yk}-\cite{Girardi:2015rwa}. 

For a model to be 
consistent, the alignment must descend from the minimization of the scalar potential, without ad-hoc assumptions on the potential parameters. 
Widely used ingredients for this type of constructions  are:
\begin{itemize}
 \item the presence of additional scalar degrees of freedom, which are called   {\it driving} fields,
and are singlets under the gauge group; 
\item additional (perhaps cyclic) symmetries, apart from $\Gr_f$, which are necessary to forbid those Lagrangian operators which would prevent 
the desired vacuum alignment.
\end{itemize}
In SUSY frameworks, both flavons and driving fields are neede to derive the superpotential $w$ of the model. In the limit of unbroken SUSY, the minimum of the related scalar potential $V$ is  given by the derivatives of $w$ with respect to the components of the driving fields, which determine a set of equations for the components  of the flavon fields. 
A detailed account of such a procedure has been given in \cite{Altarelli:2005yx}, to which we refer the interested reader.
Here we limit ourselves to a simple representative example, extracted from \cite{deMedeirosVarzielas:2005qg}. Suppose that the SM singlet pair $(\varphi _{0},\varphi )$ is made up of a  driving $(\varphi _{0})$ and a flavon $(\varphi )$ triplet fields in such a way that terms like $\varphi _{0}\varphi$ and $\varphi _{0}\varphi^2$ are flavor invariant; thus, the most general renormalisable superpotential  is given by:
\begin{eqnarray}
w &=&M(\varphi _{0}\varphi )+g(\varphi _{0}\varphi \varphi)\,.
\end{eqnarray}%
The vacuum minimisation conditions  for the $\varphi $ field are then:
\begin{eqnarray}
\frac{\partial w}{\partial \varphi _{01}} &=&M\varphi _{1}+g\varphi
_{2}\varphi _{3}=0\,,  \notag \\
\frac{\partial w}{\partial \varphi _{02}} &=&M\varphi _{2}+g\varphi
_{3}\varphi _{1}=0  \,, \\
\frac{\partial w}{\partial \varphi _{03}} &=&M\varphi _{3}+g\varphi
_{1}\varphi _{2}=0 \,,\notag
\end{eqnarray}
which are solved by:
\begin{equation}
\varphi =v\,(1,1,1),~~~~~~~v=-\frac{M}{g}.  \label{phivev}
\end{equation}
This simple case does not obviously exhaust all possible situations arising after the minimization procedure; in more complicated cases, it could happen that some of the vevs depends on unknown parameters which are not related to the  parameters appearing in $w$. This indicates that there are flat directions in the flavon potential, as one could check 
by analyzing the flavons and driving fields mass spectrum  in the SUSY limit.
SUSY breaking effects and radiative corrections are eventually important to give mass to the modes associated to these flat directions.\\
The presence of driving fields is not a necessary condition for obtaining the correct vacuum alignment. 
While this implies to deal with longer and more complicated potentials \cite{Grimus:2005mu}-\cite{Ferreira:2012ri},
one can avoid intricated calculations  formulating flavor models in extra dimensions where the scalar fields 
live in the bulk of the higher-dimensional space \cite{Kobayashi:2008ih}. The vacuum alignment is then achieved by the boundary
conditions of the scalar fields and the physics at low energy is described by
massless zero modes which break the 
flavor symmetries \cite{Burrows:2010wz}.

\subsection{Typical discrete patterns}
The use of discrete symmetries was first suggested to explain a simplified form of the neutrino mass matrix called {\it Tri-Bi-Maximal} mixing (TBM) \cite{Harrison:2002er}-\cite{Harrison:2003aw}:
\begin{eqnarray}
U_{\mathrm{TB}} =
\left( \begin{array}{ccc}
\sqrt{\frac{2}{3}} & \frac{1}{\sqrt{3}} & 0 \\
-\frac{1}{\sqrt{6}}  & \frac{1}{\sqrt{3}} & \frac{1}{\sqrt{2}} \\
\frac{1}{\sqrt{6}}  & -\frac{1}{\sqrt{3}} & \frac{1}{\sqrt{2}}
\end{array}
\right)\,,
\label{TB}
\end{eqnarray}
which implies $s^2_{12}=1/3$, $s^2_{23}=1/2$ and $s_{13}=0$.  In this case the matrix $m_\nu$ takes the form:
\begin{equation}
m_\nu=\left(\begin{array}{ccc}
x&y&y\\
y&x+v&y-v\\
y&y-v&x+v\end{array}\right)~~~,
\label{gl21}
\end{equation}
($x,y$ and $v$ are complex numbers) which can also parametrized as:
\begin{equation}
m_{\nu}=  m_1\Phi_1 \Phi_1^T + m_2\Phi_2 \Phi_2^T + m_3\Phi_3 \Phi_3^T\,, 
\label{1k1}
\end{equation}
where
\begin{equation}
\Phi_1^T=\frac{1}{\sqrt{6}}(2,-1,-1)~~~,~~~~~
\Phi_2^T=\frac{1}{\sqrt{3}}(1,1,1)~~~,~~~~~\Phi_3^T=\frac{1}{\sqrt{2}}(0,-1,1)\,,
\label{4k1}
\end{equation}
are the respective columns of $U_{TB}$ and $m_i$ are the neutrino mass eigenvalues given by the simple expressions $m_1=x-y$, $m_2=x+2y$ and $m_3=x-y+2v$ \cite{Altarelli:2010gt}.

Notice that, in the basis where charged leptons are diagonal, the mass matrix for TBM mixing is the most general matrix which is invariant under the so-called 2-3
(or $\mu-\tau$) symmetry \cite{Fukuyama:1997ky,Fukuyama:2017qxb} under which 
\bea
m_\nu=A_{23}m_\nu A_{23}~~\label{inv.1}\,,
\label{mmutau}
\eea
where $A_{23}$ is given by:
\be
A_{23}=\left(
\begin{array}{ccc}
1&0&0\\
0&0&1\\
0&1&0
\end{array}
\right)\,,
\label{Amutau}
\ee
and, in addition, under the action of a unitary symmetric matrix $S_{TB}$ which commutes with  $A_{23}$ :
\bea
m_\nu=S_{TB}m_\nu S_{TB}\,,
\label{inv}
\eea
where $S_{TB}$ is given by:
\bea
\label{trep}
S_{TB}&=\dd\frac{1}{3} \left(\begin{array}{ccc}
-1&2&2\\
2&-1&2\\
2&2&-1\end{array}\right)~~~.
\eea
In practice, the matrices $A_{23}$ and $S_{TB}$ realize the action of $Z\in \Gr_\nu$. 

For bimaximal (BM) mixing \cite{Barger:1998ta},  instead, 
we have $s^2_{12}=s^2_{23}=1/2$ and accordingly:
\begin{eqnarray}
U_{\mathrm{BM}} =
\left( \begin{array}{ccc}
\frac{1}{\sqrt{2}} & \frac{1}{\sqrt{2}} & 0\\
-\frac{1}{2}  & \frac{1}{2} & \frac{1}{\sqrt{2}} \\
\frac{1}{2}  & -\frac{1}{2} & \frac{1}{\sqrt{2}} 
\end{array}
\right)\,.
\label{BM}
\end{eqnarray}
The respective   mass matrix   is of the form:
\begin{equation}
m_{\nu}=\left(\begin{array}{ccc}
x&y&y\\
y&z&x-z\\
y&x-z&z\end{array}\right)\;,
\label{gl2}
\end{equation}
that is
\begin{equation}
m_{\nu}=  m_1\Phi_1 \Phi_1^T + m_2\Phi_2 \Phi_2^T + m_3\Phi_3 \Phi_3^T~~~, 
\label{1k}
\end{equation}
where
\be
\Phi_1^T=\frac{1}{2}(\sqrt{2},1,1)~~~,~~~~~
\Phi_2^T=\frac{1}{2}(-\sqrt{2},1,1)~~~,~~~~~\Phi_3^T=\frac{1}{\sqrt{2}}(0,-1,1)\,.
\label{4k}
\ee
The resulting matrix is characterized by the  invariance under the action of $A_{23}$ and also under the application of the real, unitary and
symmetric matrix $S_{BM}$ of the form
\be
m_{\nu}= S_{BM} m_{\nu}S_{BM}\,,
\label{invS}
\ee
with $S_{BM}$ given by:
\be
S_{BM}=\left(
    \begin{array}{ccc}
      0 & -\dd\frac{1}{\sqrt{2}} & -\dd\frac{1}{\sqrt{2}}  \\
      -\dd\frac{1}{\sqrt{2}} & \dd\frac{1}{2} & -\dd\frac{1}{2} \\
      -\dd\frac{1}{\sqrt{2}} & -\dd\frac{1}{2} & \dd\frac{1}{2} \\
    \end{array}
  \right)~~~.
  \label{matS}
\ee
In this case, are the matrices $A_{23}$ and $S_{BM}$ that realize the action of $Z\in \Gr_\nu$ on the neutrino mass matrix. 

Other examples of special patterns can be found in the literature; among them, a vast production has been devoted to the Golden Ratio mixing (GR), of which two slightly different versions have attracted  much attention:
in one of them \cite{Datta:2003qg,Kajiyama:2007gx,Everett:2008et,Feruglio:2011qq}
the solar angle is given by
$\tan \theta_{12}=1/\phi$, where $\phi = (1+\sqrt{5})/2$ is the golden ratio,
which implies $\theta_{12}=31.7^\circ$; in the other one, suggested in \cite{Rodejohann:2008ir},
$\cos \theta_{12} =\phi/2$ and $\theta_{12}=36^\circ$.

\begin{figure}[h!]
	\centering
	\includegraphics[scale=.4]{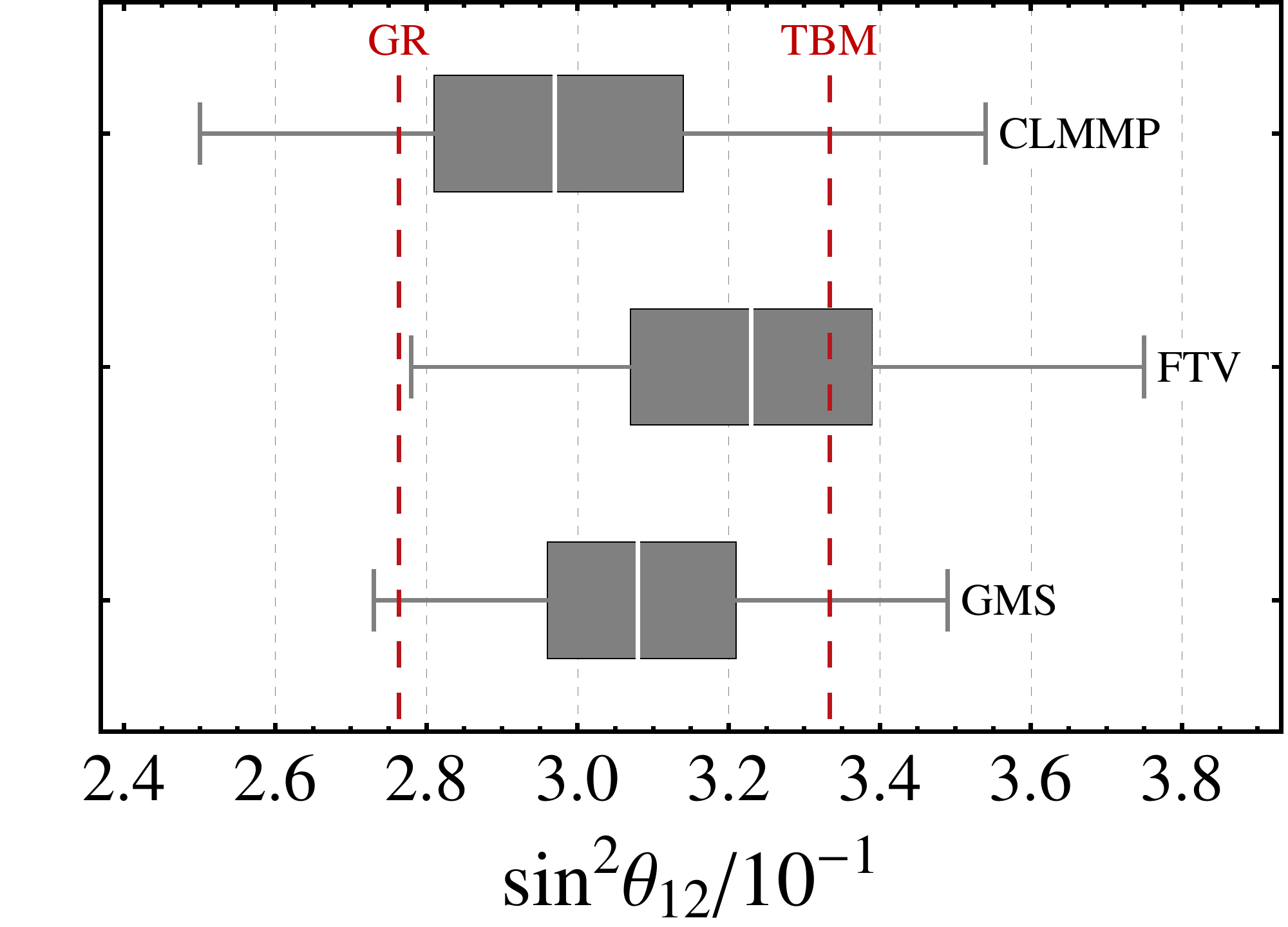}
	\caption{\it Predictions for $\sin^2\theta_{12}$ for GR and TBM mixing patterns (red dashed lines); the box charts represent the value of the global fits (for NO only since the allowed region is the same for both orderings) performed in Ref. \cite{Capozzi:2016rtj} (labeled as CLMMP), Ref. \cite{Forero:2014bxa} (labeled as FTV) and Ref. \cite{Gonzalez-Garcia:2014bfa} (labeled as GMS). The white vertical lines inside the boxes are the best fit values, the grey boxes the $1\sigma$ confidence regions and the grey lines the $3\sigma$ allowed regions.}
	\label{fig:GR_TBM_angles}
\end{figure}

Since these special patterns mainly differ for the value of the solar angle, we report in Fig.	\ref{fig:GR_TBM_angles} the predictions for $\sin^2\theta_{12}$ of GR and TBM and compare them with three different fit results coming from \cite{Capozzi:2016rtj} (labeled as CLMMP), \cite{Forero:2014bxa} (labeled as FTV) and Ref. \cite{Gonzalez-Garcia:2014bfa}  (labeled as GMS). See the caption for more details.

The neutrino mass matrices analyzed so far have been derived in the basis where charged leptons are diagonal; then one can ask which features the matrix 
$Q$ of eq.(\ref{mass_invariance_conditions}) must have in order to maintain the hermitian product $M_\ell^\dagger \, M_\ell$ diagonal; observing that the most general diagonal $M_\ell^\dagger \, M_\ell$ is left invariant under the action of a diagonal phase matrix with 3 different phase factors, one can easily see that 
if $Q^n=1$ then the matrix $Q$ generates a cyclic group $Z_n$.
Examples for $n=3$ and $n=4$ are the following:
\bea
\label{ta4}
Q_{TB}&=&\left(\begin{array}{ccc}
1&0&0\cr
0&\omega&0\cr
0&0&\omega^2 
\end{array}\right)\,,\quad \omega^3=1 \nn \\ && \\ \nn
Q_{BM}&=&\left(
    \begin{array}{ccc}
      -1 & 0 & 0 \\
      0 & -i & 0 \\
      0 & 0 & i \\
    \end{array}
  \right)\,.
\eea
We stress again that  a realistic flavor model that reproduces all experimental features of neutrino masses and mixing can be realized from a theory invariant under the spontaneously broken symmetry described by $\Gr_f$ which, in turn, must  contain at least the $S$ and $Q$ transformations. These generate the subgroups $\Gr_\nu$ and $\Gr_\ell$, respectively. 
The breaking of $\Gr_f$ must be arranged in such a way that it is broken down to $\Gr_\nu$ in the neutrino mass sector and  to $\Gr_\ell$ in the charged lepton mass sector. 
In some cases also the symmetry under $A_{23}$ is part of $\Gr_\ell$ and then must be preserved in the neutrino sector or it can arise as a consequence of the breaking of $\Gr_\ell$.

Notice that it is not strictly necessary to deal with diagonal charged leptons because the special patterns analyzed so far can be considered as a good first approximation of the data and suitable corrections, for example coming explicitly from the charged leptons, must be taken into account \cite{Petcov:2004rk,Altarelli:2004jb,Meloni:2010cj}. 
 
Many discrete groups with the previous properties have been studied and their potentialities to describe neutrino masses and mixings scrutinized in detail. 
Just to give some examples, the groups $A_4$, $S_4$ and $T'$ are commonly utilized to generate TBM mixing (see, for example, Refs.\cite{Altarelli:2005yx,Ma:2005qf,He:2006dk,Chen:2009um,Altarelli:2009kr}, 
\cite{Bazzocchi:2009pv,Grimus:2009pg} and \cite{Aranda:2007dp,Ding:2008rj,Frampton:2008bz});
the group $S_4$ can also be used to generate BM mixing \cite{Barger:1998ta,Mohapatra:1998ka, Altarelli:2009gn}; $A_5$ can be utilized to generate GR mixing \cite{Datta:2003qg,Kajiyama:2007gx,Everett:2008et,Feruglio:2011qq} and the groups $D_{10}$ and $D_{12}$ can lead to another type of GR \cite{Rodejohann:2008ir, Adulpravitchai:2009bg} 
and to hexagonal mixing \cite{Albright:2010ap,Kim:2010zub}.
Excellent reviews in this sector can be found, for instance, in Refs. \cite{King:2013eh},  \cite{Altarelli:2010gt}, \cite{Ishimori:2010au} and \cite{Grimus:2011fk}. 

\subsection{TBM and BM from discrete symmetries}
\label{sec:tbm_bm_discrete}
To make a direct connection with the procedure outlined in Sect.\ref{sec:lepton_mixing}, we study here two examples on how to get the TBM and BM patterns
from $\Gr_f = S_4$. This is the permutation group of order four, it has $4! = 24$ elements and it is isomorphic to the symmetry group of the cube. The algebra contains two generators, $S$ and $T$, that satisfy the condition $S^2 = T^4 = (ST)^3 = 1$.
\begin{table}[h!]
	\begin{center}
		\begin{tabular}{c c c c c c}
			\toprule
			\toprule
			$S_4$ & $C_1$  &$3C_2^{[2]}$ & $6C_3^{[2]}$ & $6C_4^{[4]}$ & $8C_5^{[3]}$\\ 
			\midrule 
			$\chi^{[\mathbf{1}]}$& 1 & 1 & 1 & 1 & 1\\
			$\chi^{[\mathbf{1'}]}$&1 & 1 & -1 & -1 & 1\\
			$\chi^{[\mathbf{2}]}$& 2 & 2 & 0 & 0 & -1\\
			$\chi^{[\mathbf{3}]}$& 3 & -1 & 1 & -1 & 0\\
			$\chi^{[\mathbf{3'}]}$& 3 & -1 & -1 & 1 & 0\\
			\bottomrule
			\bottomrule
		\end{tabular}
		\caption{\it Characters of the $S_4$ group.}
		\label{tab:chaS4}
	\end{center}
\end{table}
The group contains five irreducible representations: two singlets $\mbf{1}$ and $\mbf{1'}$, one doublet $\mbf{2}$ and two triplets $\mbf{3}$ and $\mbf{3'}$. The (non trivial) tensor products are
\begin{subequations}
	\begin{gather}
	\mbf{1'} \otimes \mbf{1'} = \mbf{1}\nn\nline
	\mbf{1'} \otimes \mbf{2} = \mbf{2}\nn \nline
	\mbf{1'} \otimes \mbf{3} = \mbf{3'} \nn \nline
	\mbf{1'} \otimes \mbf{3'} = \mbf{3} \nn \nline
	\mbf{2} \otimes \mbf{2} = \mbf{1}_s \oplus \mbf{2}_s \oplus \mbf{1'}_a \nn \nline
	\mbf{2} \otimes \mbf{3} = \mbf{2} \otimes \mbf{3'} =\mbf{3} \oplus \mbf{3'}\nn\nline
	\mbf{3} \otimes \mbf{3} = \mbf{3'} \otimes \mbf{3'} = \mbf{1}_s \oplus \mbf{2}_s\oplus\mbf{3'}_s \oplus \mbf{3}_a\nn\nline
	\mbf{3} \otimes \mbf{3'} =\nn \mbf{1'}\oplus\mbf{2}\oplus\mbf{3}\oplus\mbf{3'}\,,
	\end{gather}
\end{subequations}
where the subscript $s$ ($a$) denotes symmetric (antisymmetric) combinations. The $S_4$ elements can be classified by the order $h$ of each element, where $\omega^h = e$ (see Tab.\ref{tab:chaS4} where the five conjugacy classes and their characters are summarized. As expected, we have $1 + 3+ 6+6+8=24$ elements in each class and the superscript indicates the order of each element in the conjugacy classes). 
A possible choice for the three dimensional generators is
\begin{align}
\label{S4_generators_rep3}
S = \dfrac{1}{{2}}\begin{pmatrix}
0 & {\sqrt{2}} & {\sqrt{2}} \\
{\sqrt{2}} & -1 & 1 \\
{\sqrt{2}} & 1 & -1 
\end{pmatrix}
\qquad
T = \begin{pmatrix}
1 & 0 & 0\\
0 & e^{i\pi/2} & 0\\
0 & 0 & e^{i3\pi/2}
\end{pmatrix}\,.
\end{align}
The group $S_4$ contains another three dimensional representation, whose generators are related to those in \eqref{S4_generators_rep3} through $\{S, T\} \to \{-S, -T\}$. The abelian subgroups of $S_4$ are four Klein groups $V$, four $Z_3$ groups and three different $Z_4$. These are summarized in Tab. \ref{tab:S4sub}.

\begin{table}[h]
	\begin{center}
		\begin{tabular}{c c  c c c c c c}
			\toprule
			\toprule
			\multicolumn{2}{c}{$Z_4$} & \multicolumn{2}{c}{$Z_3$}& \multicolumn{2}{c}{$V$} \\
			\rowcolor{white}
			
			\bf Algebra & \bf Generators  & \bf Algebra & \bf Generators & \bf Algebra & \bf Generators \\ 
			\midrule 
			$Q_1$ & $T$   	&  $C_1$ & $ST$ 	   &$K_1$ & $\{T^2, ST^2S\}$ 	       \\
			$Q_2$ & $T^2S$ 	& $C_2$ & $TS$ & $K_2$ & $\{S, T^2ST^2\}$      \\
			$Q_3$ & $STS$ 		&$C_3$ & $T^2ST$ & $K_3$ & $\{T^2, ST^2ST\}$   \\
			& 		  	&$C_4$ & $TST^2$ & $K_4$ & $\{ST^2S, T^3ST\}$ \\
			\bottomrule
			\bottomrule
		\end{tabular}
		\caption{\it Possible independent algebras of $S_4$ subgroups (same classification as the one adopted in Ref. \cite{deAdelhartToorop:2011re}).}
		\label{tab:S4sub}
	\end{center}
\end{table}

The patterns of interest can be obtained using the following choices of subgroups:
\subsubsection*{$\bullet$ $\Gr_\ell = Z_3$ and $\Gr_\nu = V$}
These subgroups are useful to reproduce the TBM only. We assume $C_3 \in Z_3$ and $K_1 \in V$ as representative algebra. The absolute value of the PMNS matrix is therefore given by:
\begin{align}
\label{TBM_matrix}
\| \UPMNS \| = U_{\mathrm{TBM}} = \frac{1}{\sqrt{6}}\begin{pmatrix}
2 & \sqrt{2} & 0\\
1 & \sqrt{2} & \sqrt{3}\\
1 & \sqrt{2} & \sqrt{3}
\end{pmatrix}\,.
\end{align}
Notice that the Jarlskog invariant $\JCP$ \cite{Jarlskog:1985ht}, defined as:
\begin{eqnarray}
\label{JCP_definition}
\JCP &\equiv&  \Im{\bigg[(\UPMNS)_{11} (\UPMNS)^*_{13}(\UPMNS)^*_{31}(\UPMNS)_{33}\bigg]} =\nn \\ && \frac{1}{8} \sin 2\theta_{12} \sin 2\theta_{23} \sin 2\theta_{13}\cos\theta_{13} \sin\delta\,,
\end{eqnarray}
is zero. To obtain a realistic mixing pattern with $\thr \sim 9^\circ$ we need to include large corrections.

\subsubsection*{$\bullet$ $\Gr_\ell = Z_4$ and $\Gr_\nu = V$}
In this case only the BM pattern is possible;
therefore both $\theta_{12}$ and $\tha$ are maximal.  Next to leading order corrections of roughly the same order of magnitude of the Cabibbo angle are needed to reproduce the data as discussed, for instance, in Ref. \cite{Altarelli:2009gn}.

\subsubsection*{$\bullet$ $\Gr_\ell = V$ and $\Gr_\nu = V$}
This case, discussed in Ref. \cite{Lam:2011ag}, produces a BM mixing pattern. A representative choice for the subalgebras for $\Gr_\ell$ is $K_1$ and for $\Gr_\nu$ is $K_2$.

\subsection{Other LO patterns}
The fact that the value of the reactor angle   is non-zero with high accuracy opens the possibility to use discrete symmetries to enforce the LO leptonic mixing patterns to  structures  where
$\theta_{13}$ is different from zero from the beginning. The various realizations all differ by the  amount of the NLO needed to reconcile the theoretical predictions with the experimental data.
Some of the new patterns, that have been obtained and studied in specific model realizations, are the following:
\begin{itemize}
\item the Trimaximal mixing \cite{Grimus:2008tt}, which referes to schemes where  the first or the second column is the same as the corresponding one of TB matrix \cite{Albright:2010ap,He:2006qd,Antusch:2011ic}. In both cases, the good TB prediction of $\theta_{12} \sim 35^\circ$ is maintained and $\theta_{13}$ is always different from zero.
 \item the Tri-Permuting  (TP) mixing  matrix, introduced in \cite{Bazzocchi:2011ax}. The mixing  is defined by two maximal angles and a large $\theta_{13}$ according to
\bea
\sin \theta_{12}=\sin \theta_{23}= -\frac{1}{\sqrt{2}}\,, \quad \sin \theta_{13}=\frac{1}{3}\,,
\eea
which corresponds to the following mixing matrix:
\begin{equation}\label{ULO}
U_{\rm TP}\sim
\frac{1}{3}\left(
\begin{array}{ccc}
2&-2&1\\
2&1&-2\\
1&2&2
\end{array}
\right)\,.
\end{equation}
\item the Bi-trimaximal (BT) mixing, introduced in \cite{King:2012in} and corresponding to the mixing matrix:
\begin{eqnarray}
\label{Unu0}
U_{\rm{BT}}=\left(
\begin{array}{ccc}
 a_+ & \frac{1}{\sqrt{3}} & a_- \\
 -\frac{1}{\sqrt{3}} & \frac{1}{\sqrt{3}} & \frac{1}{\sqrt{3}} \\
a_- & -\frac{1}{\sqrt{3}} & a_+
\end{array}
\right)\,,
\end{eqnarray}
where $a_{\pm}=(1\pm \frac{1}{\sqrt{3}})/2$, and 
leads to the following predictions: 
\begin{eqnarray}
\label{delta96pred2}
\begin{array}{l}
\sin \theta_{12} =\sin \theta_{23}= \sqrt{\frac{8-2\sqrt{3}}{13}}\approx 0.591 
\ \  \ \ (\theta_{12}=\theta_{23}\approx 36.2^{\circ}), \\
\sin \theta_{13}= a_-\approx 0.211 
\ \  \ \ (\theta_{13}\approx 12.2^{\circ}).
\end{array}
\end{eqnarray}

\end{itemize}

\subsection{Discrete symmetries and invariance under CP}
\label{CPsymm}
Let us now enlarge the symmetry content of the theory assuming, in addition to the invariance under the discrete group, also invariance under $CP$ \cite{Feruglio:2012cw, Holthausen:2012dk,Chen:2014tpa}. 

As in Sec. \ref{sec:lepton_mixing}, we consider that the residual symmetry in the charged sector $\Gr_\ell$ is a cyclic group $Z_n, n\ge 3$, or the product $Z_{2} \otimes Z_{2}$.
 Under the action of $CP$, a generic field $\Phi$ transforms as \cite{Ecker:1983hz, Ecker:1987qp, Neufeld:1987wa}:
\begin{align}
\label{Xfield}
 \Phi(x) \longrightarrow \Phi'(x) = X \Phi^\star(x_{CP})\,,
\end{align}
where $X$ is the representations of the $CP$ operator in field space and $x_{CP}$ is the space-time coordinate transformed under the usual $CP$ transformation $x \to x_{CP} = (x^0, - \mathbf{x})$. The invariance of the field under $\Gr_f$ is expressed as:
\begin{align}
\label{Afield}
\Phi(x) \longrightarrow \Phi'(x) = A \Phi(x)\,,
\end{align}
where $A$ is an element of a non-abelian discrete symmetry group. $X$ can be chosen as a constant unitary symmetric matrix\footnote{The requirement that $X$  is a symmetric
matrix has been shown in \cite{Feruglio:2012cw} to be a necessary condition, otherwise the neutrino mass spectrum
would be partially degenerate.}:
\begin{align}
\label{Xunitary}
 XX^\dagger = XX^\star = 1\,,
\end{align}
in such a way that the square of the $CP$ transformation is the identity, $X^2 = 1$. The action of $X$ on the mass matrices, before the symmetry breaking, is given by
\begin{subequations}
\begin{gather}
 X^\star M_\ell^\dagger M_\ell X = (M_\ell^\dagger M_\ell)^\star\\
 \label{Xmajorana}
 X M_\nu X = M_\nu^\star\,,
\end{gather}
\end{subequations}
if neutrinos are Majorana particles. If instead neutrinos are Dirac particles, \eqref{Xmajorana} has to be modified to
\begin{align}
 X^\star M_\nu^\dagger M_\nu X &= (M_\nu^\dagger M_\nu)^\star\,.
\end{align}
The fact that the theory is invariant under  the flavour symmetry group $\Gr_f$ requires that for the generators of the group $A$ the representations $X$ in the field space must satisfy the following relation:
\begin{align}
\label{Xconditionsgenerator}
 (X^{-1} A X)^\star = A'\qquad A, A' \in \{\Gr_f\}\,,
\end{align}
where in general $A \not= A'$.  Notice that if $X$ is a solution of \eqref{Xunitary} and \eqref{Xconditionsgenerator} also $e^{i \rho}X$, with $\rho$ being an arbitrary phase, is a solution.  

Let us now specify this framework to the case where the residual symmetry $\Gr_\nu$ is $Z_2 \otimes CP$, with $Z_2$ contained in the flavour group;  the matrix $Z$ representing the generator of the former symmetry and the CP transformation $X$ have to fulfill the constraint
\begin{align}
\label{conditionZX}
 XZ^\star - ZX = 0\,,
\end{align}
which is invariant under \eqref{Xconditionsgenerator}. 
In the neutrino sector, the light neutrino mass matrix satisfies both relations:
\begin{subequations}
\label{neutrino_mass_conditions}
\begin{align}
 Z^T M_\nu Z &= M_\nu \nline
 X M_\nu X &= M_\nu^\star\,.
\end{align}
\end{subequations}
Notice that it is always possible to choose a basis where
\begin{align}
\label{Z_c_equation}
X = \Omega \Omega^T \qquad  Z_c = \Omega^\dagger Z \Omega \qquad Z_c = \diag\Big\{(-1)^{z_1},(-1)^{z_2},(-1)^{z_3} \Big\}\,,
\end{align}
with $z_i = 0, 1$.  
Since $Z$ generates a $Z_2$ symmetry, two of the three
parameters $z_i$ have to coincide and the combination $\Omega^T M_\nu \Omega$ is constrained to be block-diagonal and real. Thus this matrix can be diagonalized using a rotation $R_{\ij}(\theta)$ in the $ij$-plane of degenerate eigenvalues of $Z$, where $\theta$ is an unconstrained parameter that can be fixed to describe the neutrino mixing parameters. 
The positiveness of the light neutrino masses  is ensured by the diagonal matrix $K_\nu$ with elements equal to $\pm 1$ or $\pm i$. In this way the matrix $M_\nu$ can be diagonalized with unitary matrix defined as
\begin{align}
 U_\nu \equiv \Omega R_{ij}(\theta) K_\nu.
\end{align}
The mass spectrum is not fixed and thus permutations of columns are admitted.
The inclusion of the charged leptons into the game proceeds 
as discussed in Sec. \ref{sec:lepton_mixing}. So, called $U_\ell$ the matrix diagonalizing $M_\ell^\dagger M_\ell$, the full $\UPMNS$ is given by:
\begin{align}
\label{PMNS_matrix_definition_CP}
 \UPMNS \equiv U_\ell^\dagger U_\nu = U_\ell \Omega R_{ij}(\theta) K_\nu\,,
\end{align}
up to permutations of rows and columns. 
To give an explicit example \cite{DiIura:2015kfa}, let us assume that 
$U_\ell=1$ and  take $\Omega$ to be
\begin{equation}
\label{GeZ5Omega1}
\Omega= \frac{1}{\sqrt{2}} \, \left(
\begin{array}{ccc}
 \sqrt{2} \, \cos\varphi & -\sqrt{2} \, i \, \sin\varphi & 0\\
 \sin\varphi & i \, \cos\varphi & -1\\
 \sin\varphi &  i \, \cos\varphi & 1
\end{array}
\right) \; ;
\end{equation}
this matrix  fulfills (\ref{Z_c_equation}) for $Z$ and $X$ chosen as $(Z,X)$ = $(T^2 S T^3 S T^2, S X_0)$, with $X_0 \equiv A_{23}$. 
Since $z_1$ and $z_3$ of the diagonal combination $\Omega^\dagger \, Z  \, \Omega$ are
equal, the indices $ij$ of the rotation matrix $R_{ij} (\theta)$ in (\ref{PMNS_matrix_definition_CP}) are $\{i,j\}=\{1,3\}$. Thus, the PMNS mixing matrix simply reads 
\begin{equation}
\label{UPMNS1}
U_{PMNS}= \Omega \, R_{13} (\theta) \, K_\nu \; .
\end{equation}
Extracting the mixing angles from (\ref{UPMNS1})  we find:
\begin{eqnarray}\nonumber
&&\sin^2 \theta_{12} = \frac{2}{2 + (3+\sqrt{5}) \, \cos^2 \theta} \;\;\; , \;\;\; \sin^2\theta_{13}= \frac{1}{10} \, \left( 5+\sqrt{5} \right) \, \sin^2 \theta \; ,
\\  \label{anglesCaseI}
&&\sin^2 \theta_{23}= \frac 12 - \frac{\sqrt{2 \, (5+\sqrt{5})} \, \sin 2\theta}{7+\sqrt{5} + (3+\sqrt{5}) \, \cos 2 \theta}  \; ,
\end{eqnarray}
which also call for an exact sum rule among the solar and the reactor mixing angles:
\begin{equation}
\label{GeZ5sumrule1}
\sin^2 \theta_{12}= \frac{\sin^2\varphi}{1-\sin^2 \theta_{13}} \approx \frac{0.276}{1-\sin^2 \theta_{13}}  \; .
\end{equation}
Using for $\sin^2\theta_{13}$ its best fit value $(\sin^2 \theta_{13})^\mathrm{bf}=0.0217$, we find for the solar mixing angle $\sin^2\theta_{12} \approx 0.282$
which is within its $3 \, \sigma$ range, see Tab.(\ref{tab:oscillation_exp}).

Models that explore the predictability of the $CP$ symmetry in conjunction with non-abelian discrete symmetries have been massively explored in the very recent years; for example, 
the interplay between 
$S_4$ and CP has been studied, among others, in \cite{Mohapatra:2012tb,Feruglio:2013hia,Luhn:2013vna,Penedo:2017vtf}, while the role of $A_5$ has been elucidated in \cite{Li:2015jxa,Ballett:2015wia,Turner:2015uta} and that of several $\Delta$ groups in \cite{deMedeirosVarzielas:2011zw, Bhattacharyya:2012pi,Ma:2013xqa,Hagedorn:2014wha,Ding:2015rwa}.

\subsection{The use of abelian symmetries}

Let us now investigate the possibility to construct SUSY models where the only flavor symmetry is a continuous $U(1)$~\cite{Froggatt:1978nt}; 
thus the following procedure can be used: 
\begin{itemize}
\item[-] given that the flavour symmetry acts horizontally on leptons, the related charges can be written as $e^c\sim(n_1^\R,n_2^\R,0)$ for the $SU(2)_\L$ lepton singlets and as $L\sim(n_1^\L,n_2^\L,0)$ for the lepton doublets. Since only charge differences  impact the mass hierarchies and the mixing angles, the third lepton charges can be set to zero and one can safely assume a charge ordering as $n_1^\R>n_2^\R>0$. To prevent flavour-violating Higgs couplings, the Higgs fields $H_{u,d}$ are not charged.

\item[-] Once we have assigned  $U(1)$ charges to leptons, the Yukawa terms are no longer invariant under the action of the flavour symmetry and new scalar fields $\theta$ must be introduced that transforms non-trivially under $U(1)$, with charge $n_\theta$. Thus, the Yukawa part of the Lagrangian is as follows:
\be
\begin{split}
\mathcal{L}_Y =&\,(Y_e)_{ij}\,L_i\,H_d\, e^c_j \left(\dfrac{\theta}{\Lambda}\right)^{p_e}
+(Y_\nu)_{ij}\,\dfrac{L_i L_j H_u H_u}{\Lambda_\L}\left(\dfrac{\theta}{\Lambda}\right)^{p_\nu} +\text{H.c.}
\end{split}
\label{Yukawas}
\ee
where $\Lambda$ is the cut-off of the effective flavour theory and $\Lambda_\L$ the scale of the lepton number violation, in principle distinct from $\Lambda$. 
Here $(Y_e)_{ij}$ and $(Y_\nu)_{ij}$ are free complex parameters with modulus of ${\cal O}(1)$ while $p_e$ and $p_\nu$ are appropriate powers of the ratio $\theta/\Lambda$ needed to compensate the $U(1)$ charges for each Yukawa term. 
Without loss of generality, we can fix $n_\theta=-1$; consequently,  $n_1^{L,R}, n_2^{L,R} > 0$ for the Lagrangian expansion to make sense. 
For the neutrino masses we consider that they are described  by the effective Weinberg operator, while the extension to  see-saw mechanisms is straightforward.

\item[-] Once the flavour and electroweak symmetries are broken by the vevs of the flavon and the Higgs fields, the  mass matrices arise, with entries proportional to the expanding parameter 
$
\ep\equiv\dfrac{\mean{\theta}}{\Lambda}<1$.
\end{itemize}

The lepton charges assignments reported in Tab.(\ref{tab-models}), some of them already studied in \cite{Altarelli:2012ia}, give rise to the following mass matrices \cite{Bergstrom:2014owa}:
\begin{table}[h!]
\begin{center}
\begin{tabular}{|c|c|c|}
\hline
& & \\ [-2mm]
{Model}& $e^c$ & $L$ \\ [2mm]
\hline
& & \\ 
{Anarchy ($A$)}& (3,2,0)& (0,0,0)\\ [2mm]
{$\mu\tau$-Anarchy ($A_{\mu\tau} $)}& (3,2,0) & (1,0,0)\\ [2mm]
{Hierarchy ($H$)}& (5,3,0) & (2,1,0) \\ [2mm]
\hline
& & \\ [-2mm]
{New Anarchy ($A'$)}& (3,1,0) & (0,0,0) \\ [2mm]
{New Hierarchy ($H'$)}& (8,3,0) & (2,1,0)\\ [2mm]
\hline
\end{tabular}
\end{center}
\caption{\it Examples of charge assignment under $U(1)$. With {\it anarchy} we refer to models where no symmetry at all is acting on the neutrino sector \cite{Hall:1999sn,Haba:2000be,deGouvea:2003xe} and so the charge of the lepton doublets is vanishing.}
\label{tab-models}
\end{table}

\be
\begin{aligned}
A:\quad 
Y_e&=\left(  \begin{matrix}  \ep^3 &    \ep^2 &   1 \\  \ep^3 &    \ep^2 &   1 \\ \ep^3 &    \ep^2 &   1 \end{matrix}\right)\,,\;
Y_\nu= \left(  \begin{matrix}  1 &    1 &   1 \\  1 &   1 &  1 \\ 1 &   1 &   1 \end{matrix}\right)\,,\\
A_{\mu\tau}:\quad 
Y_e&= \left(  \begin{matrix}  \ep^4 &    \ep^3 &   \ep \\  \ep^3 &    \ep^2 &   1 \\ \ep^3 &   \ep^2 &   1 \end{matrix}    \right)\,,\;
Y_\nu= \left(  \begin{matrix}  \ep^2 &    \ep &   \ep \\  \ep &   1 &  1 \\ \ep &   1 &   1 \end{matrix} \right)\,,  \\
H:\quad 
Y_e&= \left(  \begin{matrix}  \ep^7 &    \ep^5 &   \ep^2 \\  \ep^6 &    \ep^4 &   \ep \\ \ep^5 &   \ep^3 &   1 \end{matrix} \right)\,,\;
Y_\nu= \left(  \begin{matrix}  \ep^4 &    \ep^3 &   \ep^2 \\  \ep^3 &   \ep^2 &  \ep \\ \ep^2 &   \ep &   1 \end{matrix}     \right)\,,
\end{aligned}
\label{OldModels} 
\ee
\be
\begin{aligned}
A':\quad 
Y_e&=\left(  \begin{matrix}  \ep^3 &    \ep &   1 \\  \ep^3 &    \ep &   1 \\ \ep^3 &    \ep &   1 \end{matrix}\right)\,,\;
Y_\nu= \left(  \begin{matrix}  1 &    1 &   1 \\  1 &   1 &  1 \\ 1 &   1 &   1 \end{matrix}\right)\,,\\
H':\quad 
Y_e&= \left(  \begin{matrix}  \ep^{10} &    \ep^6 &   \ep^2 \\  \ep^9 &    \ep^5 &   \ep \\ \ep^8 &   \ep^4 &   1 \end{matrix} \right) ,\;
Y_\nu= \left(  \begin{matrix}  \ep^4 &    \ep^3 &   \ep^2 \\  \ep^3 &   \ep^2 &  \ep \\ \ep^2 &   \ep &   1 \end{matrix}     \right)\,.
\end{aligned}
\label{NewModels} 
\ee

\begin{figure}[h!]
\begin{center}
\includegraphics[width=0.48\textwidth]{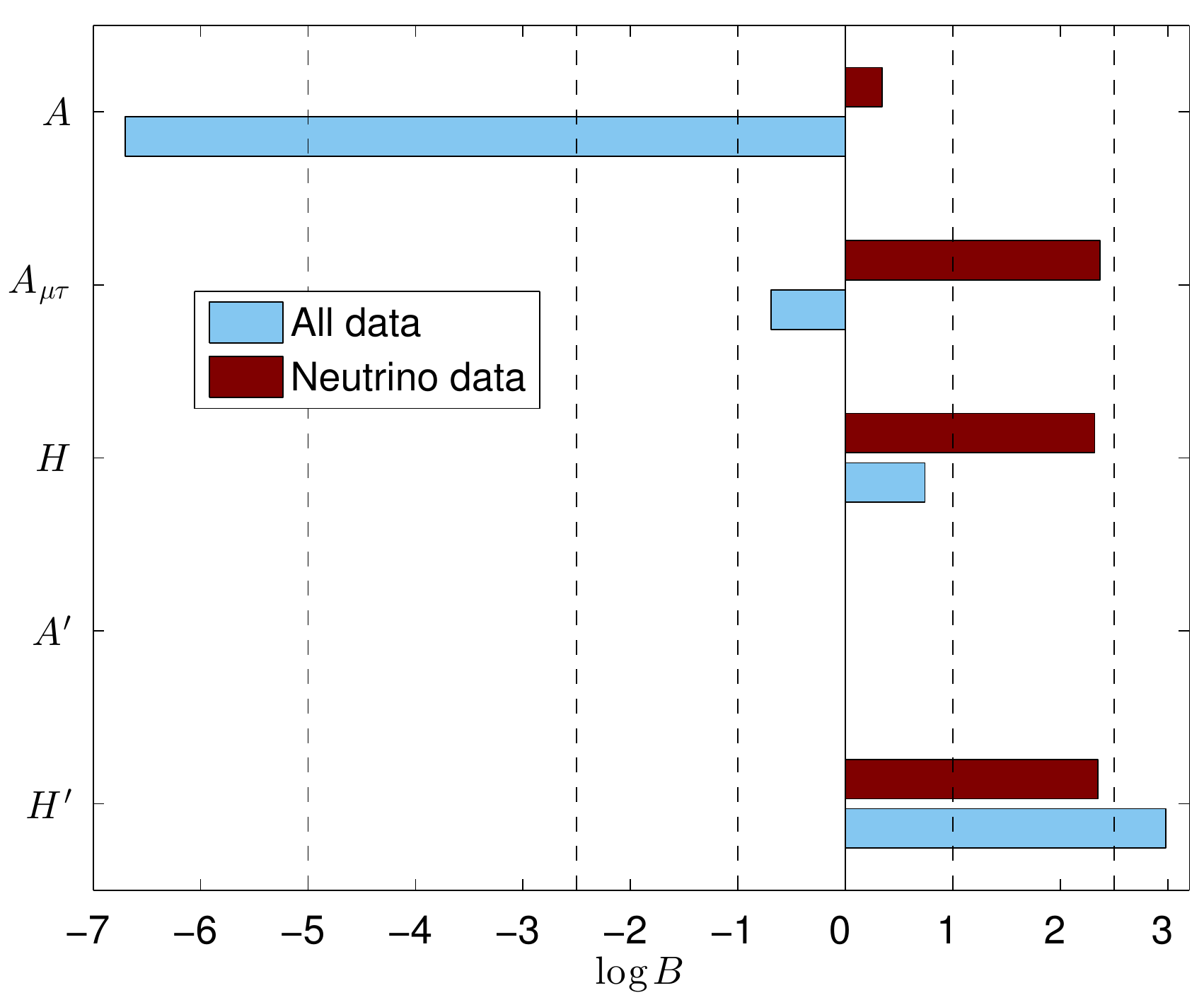}
\caption{\it Logarithms of Bayes factors with respect to the model $A'$ for the models in  Tab.(\ref{tab-models}) using only neutrino data (dark-red bars) and all data (light-blue bars). 
Positive values of $\log B$ indicate a weak evidence $(\log B=1)$, a moderate evidence $(\log B=2.5)$ and a strong evidence $(\log B=5)$ of the supposed model against the reference one $A'$. Numerical estimates on $\ep$ are not reported  but values in the range (0.1-0.2) emerged from the analysis as the most appropriate ones.}
\label{fig:Bfactors}
\end{center}
\end{figure}

As already remarked, the coefficients in front of $\ep^n$ are complex numbers with absolute values of ${\cal O}(1)$ and arbitrary phases.
Considering that $Y_\nu$ is a symmetric matrix, the total number of undetermined parameters that arise in this type of constructions
is $30$  plus the unknown value of $\ep$. In order to establish which models adapt better to the data of Tab.(\ref{tab:oscillation_exp}), one cannot use 
a $\chi^2$-based analysis because the minimum is always very close to zero for every $(Y_e,Y_\nu)$ pairs; thus, a meaningful comparison of two models is better achieved  with the help of a  
Bayesian analysis. This has been done in  \cite{Bergstrom:2014owa} and the results of the Bayes factor between all models and $A'$ are reported in Fig.(\ref{fig:Bfactors}).
The relevant features of such an analysis can be summarized as follows: when using only the neutrino data, the hierarchical models are all weakly preferred over the anarchical ones. 
When also the charged lepton data are taken into account in the analysis, the $A$ model turns out to be strongly disfavored. 
Adding in the comparison also the $H'$ and $A'$ models, the former is the best one: it is moderately better 
than $A_{\mu\tau}$ and $A'$, and weakly preferred over $H$. 

Other possibilities in the direction of using $U(1)$ rely on the fact that the $U(1)$ charges are not completely arbitrary but are determined by an underlying symmetry
of the type  $L_e-L_\mu-L_\tau$ for lepton doublets and arbitrary right-handed charges  \cite{Petcov:1982ya,Altarelli:2005pj,Meloni:2011ac}.
In the limit of exact symmetry,  the neutrino mass matrix has the following structure:
\bea m_\nu =m_0
\left(
\begin{array}{ccc}
0& 1& x\\ 1& 0&0\\ x& 0&0
\end{array}
\right)\,,
\label{mass}
\eea
which leads to a spectrum of inverted type and mixing angles as $\theta_{12} = \pi/4$, $\tan \theta_{23} = x$ (i.e. large atmospheric mixing for $x\sim {\cal O}(1)$) and $\theta_{13} = 0$.
An important limitation of such a texture is that two eigenvalues have the same absolute values and the solar mass difference cannot be reproduced.
Successful tentatives to describe also $\Delta m^2_{21}$  have been presented, for instance, in \cite{Lavoura:2000ci,Grimus:2004cj} where,
however, either the reactor angle was almost vanishing or
the solar angle was too large with respect to its current value. 
Corrections of $\cal{O}(\lambda_{\it C})$ from the charged lepton sector \cite{Petcov:2004rk,Altarelli:2004jb,Meloni:2010cj} could be invoked to properly shift 
$\theta_{12}$ from maximal mixing and $\theta_{13}$ from zero, thus allowing a sizable reactor angle,
but at the prize of a too large solar-to-atmospheric mass ratio $r$. 
A possible solution  to the previous issues was discussed in \cite{Meloni:2011ac}, where the  $U(1)$ flavour symmetry was broken by the vevs of {\it two} 
complex fields $\phi$ and $\theta$ (instead of one) of charges $Q_\phi=1$ and $Q_\theta=-1/2$. An appropriate breaking of $L_e-L_\mu-L_\tau$
in the neutrino sector assures the correct value of $r\sim \lambda_C^2$ and  preserves the leading order (LO)
prediction of large $\theta_{23}$, whereas  the necessary
deviations for the solar and reactor angles  are instead obtained from the charged lepton mass matrix with complex entries.

\section{Where GUT meets Flavor}
\label{gutflav}
\begin{figure}[h!]
\begin{center}
\includegraphics[width=0.525\textwidth]{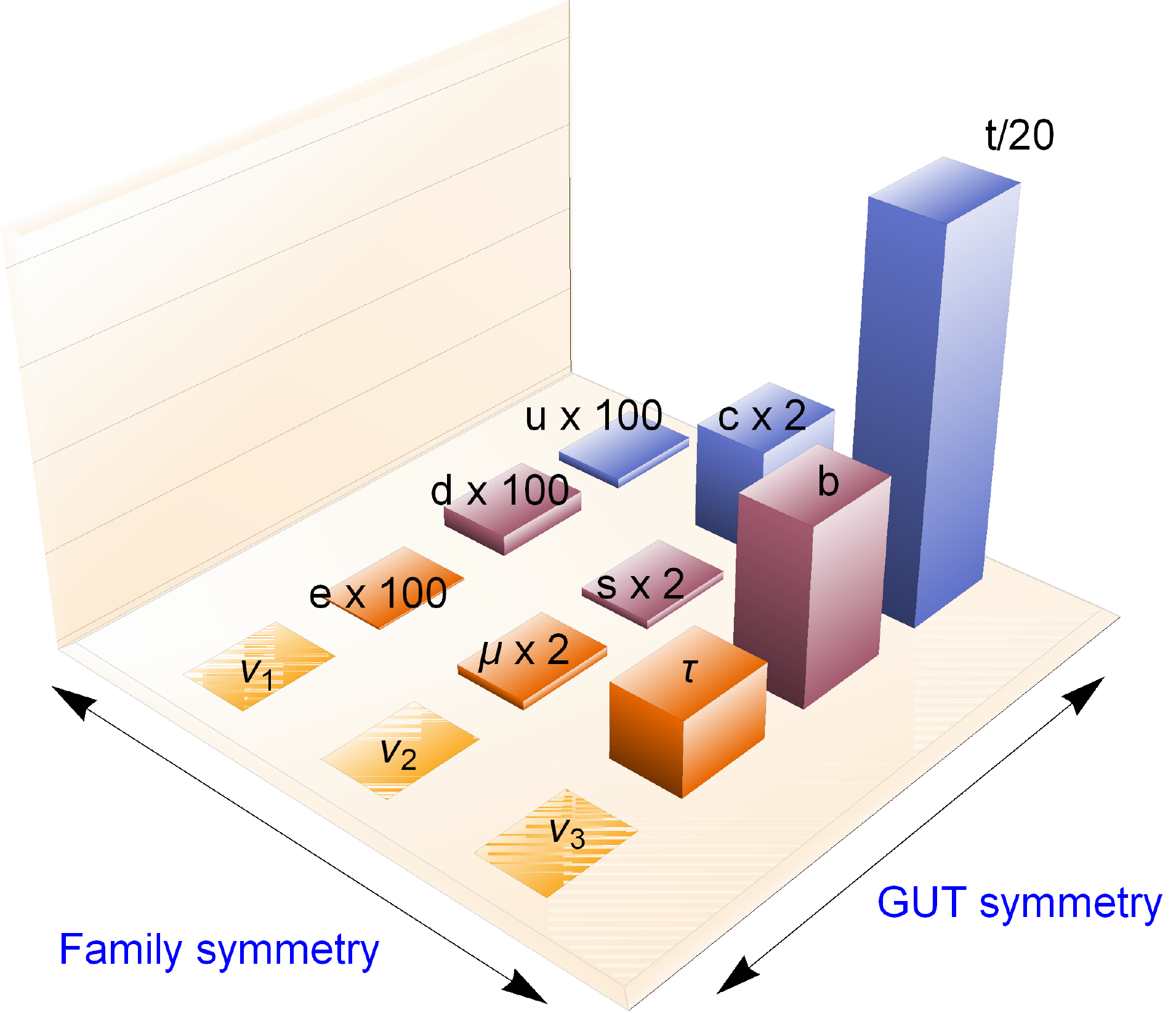}
\end{center}
\vspace*{-4mm}
    \caption{\label{interplay}\small{\it  Action of the GUT and family symmetry groups.    Given the large hierarchies, the height of the columns are not in scale and the actual values of the fermion masses have been multiplied or divided by the factors on top of each columns.}}
\vspace*{-2mm}
\end{figure}

The importance of the discovery of neutrino masses and mixing angles is
that they provide interesting information on the problem of understanding
the origin of three families of quarks and leptons and their  mixing parameters.
In this respect, as we have already outlined before, the relevance of GUT groups resides on the fact that some of the mass matrices of different fermions are related 
in a non-trivial way, see for example eq.(\ref{masses}), whereas family symmetries impose stringent constraints on the matrix elements of the same mass matrix. 
Fig.(\ref{interplay})  summarizes in concise way how GUT and family symmetries act on the observable fermions (see caption for more details).

\begin{figure}[h!]
\centerline{\includegraphics[width=8.5 cm]{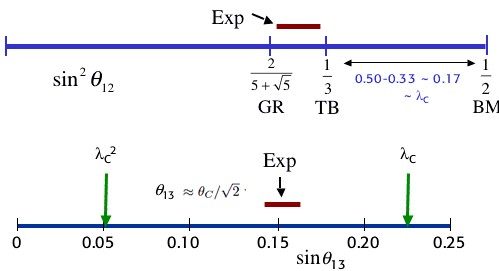}}
\caption{\it Comparison of the experimental value of the leptonic mixing angles against their exact predictions by TBM, GR and BM mixings. Figure from \cite{Altarelli:2015foa}. \label{f1}}
\end{figure}

The next obvious step is to merge these two different type of symmetries in order to construct a flavor sector with very few free parameters.
As it was the case for the special patterns of lepton mixing, also in the case with GUT one needs to identify which features of the data are really relevant for the formulation of a model. In this sense, the fact that the reactor angle $\theta_{13}$ is approximately related to the Cabibbo angle $\theta_C$ by the relation 
$\theta_{13}\sim \theta_C / \sqrt{2} $ may be a hint of a connection between leptonic and quark mixing \cite{Antusch:2005ca}. And this is not restricted to the reactor angle only. 
In fact, as shown in Fig.(\ref{f1}), the experimental value of $\sin^2{\theta_{12}}$ is related to the predictions of exact TBM or GR 
by a jump of order $\lambda_C^2$, or of order $\lambda_C$ in the case of BM. 
This idea seems to agree with the empirical observation that $\theta_{12}+\theta_C\sim \pi/4$, a relation 
known as {\it quark-lepton complementarity} \cite{Raidal:2004iw}-\cite{Frampton:2004vw}, sometimes replaced by 
$\theta_{12}+\mathcal{O}(\theta_C)\sim \pi/4$ ({\it weak complementarity}).
If we  want to realize in a complete model the previous relations, one possibility is
 to start from BM and generate universal corrections to the mixing angles of order $\lambda_C$, arriving at the following relations:
\be
\sin^2 \theta_{12} \sim \frac{1}{2} + {\cal O} (\lambda_C) \qquad \sin^2 \theta_{23} \sim \frac{1}{2} + {\cal O} (\lambda_C) \qquad
\sin \theta_{13} \sim   {\cal O} (\lambda_C)\,,
\ee
which are all in agreement with the experimental data. These corrections can be appropriately fabricated  by charged lepton rotations which differ from the identity by off-diagonal elements whose magnitude is obviously of order of the Cabibbo angle. The game becomes highly non trivial in GUT theories which demand that also masses for the quarks and the CKM matrix are reproduced at the same time.
An example based on $SU(5)$ that permits to realize the program of having
the BM structure in the neutrino sector and then to correct it by terms arising from the diagonalization of the charged lepton mass matrix is built as follows \cite{Meloni:2011fx} (but see \cite{Altarelli:2008bg} for a variant using the $A_4$ family group).
The construction is a SUSY $SU(5)$ model in 4+1 dimensions \cite{Kawamura:2000ev,Hall:2002ci} with a flavour symmetry $S_4 \otimes Z_3 \otimes U(1)_R \otimes U(1)$ \cite{Altarelli:2009gn,Meloni:2011fx}, 
where $U(1)$ is the Froggatt-Nielsen (FN) symmetry  that leads to the hierarchies of fermion masses 
and $U(1)_R$ is the usual  R-symmetry. 
The particle assignments are displayed in Tab.\ref{tab:Multiplet1} where, for the sake of simplicity, we have not reported the driving fields needed to realize the wanted symmetry breaking pattern. From the table we see that the three  $\overline{5}$ are grouped into the $S_4$ triplet $F$, while the
tenplets $T_{1,2,3}$ are assigned to the singlet of $S_4$. The breaking of the $S_4$ symmetry is ensured by  a set of  SU(5)-invariant flavon
supermultiplets, which are three triplets $\varphi_\ell, \varphi_\nu$ ($3_1$), $\chi_\ell$ ($3_2$) and 
one singlet $\xi_\nu$. The alignment in flavor space of their vevs along appropriate directions will be the source of the BM lepton mixing.  The GUT Higgs fields $H^{}_{5}$ and $H_{\overline{5}}$ are singlets under $S_4$ but equally charged under $Z_3$,
so that they are distinguished only by their SU(5) transformation properties.
The tenplets $T_1$ and $T_2$ are charged under the $U(1)$ flavour group which is spontaneously broken by the vevs of the
$\theta$ and $\theta^\prime$ fields,  both carrying $U(1)$ charges $-1$ and transforming as a singlet of $S_4$. 
\begin{table}[h!]
\small
\begin{center}
\begin{tabular}{|c||c|c|c|c|c|c|c|c|c|c|c||}
\hline
{\tt Field} & $F$ & $T_1$ & $T_2$ & $T_3$ & $H_5,H_{\overline 5}$ &   $\varphi_\nu$ & $\xi_\nu$ &  $\varphi_\ell$& $\chi_\ell$ &  
$\theta$ &  $\theta^\prime$ 
\\
\hline\hline
SU(5) & $\bar{5}$  & 10 & 10 & 10 & 5,$\overline 5$ &   1 & 1 & 1 & 1 & 1 & 1   \\
\hline
$S_4$  &  $3_1$ & 1  & 1 & 1 & 1 &   $3_1$ & 1 & $3_1$ & $3_2$ &  1 & 1      \\
\hline
$Z_3$ & $\omega$ & $\omega$ &1  &  $\omega^2$ & $\omega^2$ & 1 & 1 & $\omega$ &$\omega$ &1 &  $\omega$   \\
\hline
$U(1)_R$ & 1 & 1 & 1 & 1 & 0 & 0&  0 & 0 & 0 & 0& 0   \\
\hline
$U(1)$ & 0 & 2 & 1 & 0 & 0 &  0 & 0 & 0 & 0& -1 & -1  \\
\hline 
  & {\tt br} &  {\tt bu} & {\tt bu} & {\tt br} &  {\tt bu} &     {\tt br} &  {\tt br} &  {\tt br} & 
 {\tt br} &  {\tt br} &  {\tt br} \\
\hline 
\end{tabular}
\caption{\label{tab:Multiplet1}\it Matter and Higgs assignment of the model. The symbol ${\tt br}({\tt bu})$ indicates that the corresponding 
fields live
on the brane (bulk).}
\end{center}
\end{table}

As a result of symmetries and field assignments to the irreducible representations of $SU(5) \times S_4$, the charged lepton masses are diagonal at LO and exact BM is achieved for neutrinos.  
Higher dimension vertices in the Lagrangian, suppressed by 
powers of a large scale $\Lambda$, generate corrections to the diagonal charged leptons and to exact BM. 
We adopt the definitions:
\bea
\frac{v_{\varphi_\ell}}{\Lambda} \sim \frac{v_{\chi}}{\Lambda} 
\sim \frac{v_{\varphi_\nu}}{\Lambda} \sim \frac{v_{\xi}}{\Lambda}\sim
\frac{\langle \theta \rangle}{\Lambda} \sim \frac{\langle \theta^\prime \rangle}{\Lambda} \sim
s\sim \lambda_C\,,
\label{vsup}
\eea  
where $s=\dd\frac{1}{\sqrt{\pi R \Lambda}}$ is the volume suppression factor and $v_\phi$ are the vevs of the flavon fields listed in Tab.\ref{tab:Multiplet1}.
This simple (and democratic) choice leads to a good description of masses and mixing.
In fact, the charged lepton mass matrix turns out to be:
\be
m_e \sim \left(
\begin{array}{ccc}
a_{11}\lambda_C^5&a_{21} \lambda_C^4& a_{31}\lambda_C^2\\
a_{12} \lambda_C^4&-c\, \lambda_C^3& ...... \\
a_{13}\lambda_C^4&c \,\lambda_C^3&a_{33} \lambda_C
\end{array}
\right) \lambda_C ,
\label{melam}
\ee
where the $a_{ij}$ are generic complex coefficients  of modulus of $\mathcal{O}(1)$ not predicted by the theory.
The corresponding lepton rotation is thus:
\bea
\label{ul}
U_\ell \sim  
\left(
\begin{array}{ccc}
1  &u_{12}\lambda_C&u_{13}\lambda_C  \\
-u_{12}^* \lambda_C&  1 & 0 \\
-u_{13}^* \lambda_C   &  
-u_{12}^* u_{13}^*\lambda_C^2     & 1 \\
                     \end{array}
                   \right)\,,
\eea 
($u_{ij}$ again of  $\mathcal{O}(1)$) so that $\theta_{23}^\ell = 0$.

The neutrino masses are obtained by Weinberg operators of the form:
\bea
(FF)_1 (H_5 H_5)\,, (FF)_{3_1} H_5 H_5 \varphi_\nu \,, (FF)_{3_1} H_5 H_5 \xi_\nu\,,
\eea
which are diagonalized by exact BM,  so the mixing angles are easily derived:
\bea
\sin^2{\theta_{12}} =\frac{1}{2}- \frac{1}{\sqrt{2}}~Re(u_{12}+u_{13})\lambda_C   \qquad
\sin^2{\theta_{23}} =\frac{1}{2}+ \mathcal{O}(\lambda_C^2) \qquad
\sin{\theta_{13}} = \frac{1}{\sqrt{2}} |u_{12}-u_{13}|\lambda_C \nn\,.
\eea
We observe that  the model produces at the same time the "weak" complementarity relation and the empirical fact that 
$\sin{\theta_{13}}$ is of the same order 
than the shift of 
$\sin^2{\theta_{12}}$ from the BM value of 1/2, both of order $\lambda_C$.

It is important to stress that the  predictions of GUT models are valid at the GUT scale and, in order to compare with the experimental results, the evolution of the Yukawa matrices down to the electroweak scale must be performed \cite{Antusch:2010es,Antusch:2013wn}. Although the final values depend somehow on the details of the model, it is known that 
in the case of a quasi-degenerate neutrino mass spectrum, the renormalization group corrections to the neutrino parameters can be dramatically large \cite{Chankowski:2001mx,Kuo:2001ha}. 
However, as it has been elucidated  in \cite{Antusch:2003kp,Antusch:2005gp}, in SUSY models  small $\tan \beta$ and small neutrino
Yukawa couplings are sufficient conditions for having the corrections to the
mixing angles (and CP phases) are under control.

The requirement of having a BM mixing as a starting point is not a necessary ingredient to get a good description of fermion observables; as pointed out in \cite{Hagedorn:2012ut}, 
even from the TBM at LO one can conceive a model where the corrections to the reactor angle are large enough to meet the experimental value, maintaining at the same time the solar and atmospheric mixing at acceptable values. Also the choice of the discrete group is not restricted to $S_4$; examples where a large $\theta_{13}$ is obtained after substantial corrections from higher order operators can be found, for example, in \cite{Antusch:2013wn,Cooper:2012wf,Marzocca:2011dh,Antusch:2013kna,Antusch:2013tta,Bjorkeroth:2015ora,Antusch:2017ano}, \cite{Gehrlein:2014wda}, \cite{Meroni:2012ty} and \cite{King:2012in}, which employ the $A_4$, $A_5$, $T^\prime$ and $\Delta(96)$ groups, respectively, within an $SU(5)$ framework.


If the gauge group is enlarged to $SO(10)$, we loose the advantages of using the 
$SU(5)$-singlet right-handed neutrinos since one generation of  fermion belongs to the $\bf 16$-dimensional representation.
%
One possible strategy to separate neutrinos from the charged fermions 
 is to assume the dominance of type-II see-saw  with respect to the more usual type-I see-saw.

\begin{figure}[h!]
\begin{center}
\includegraphics[scale=.5]{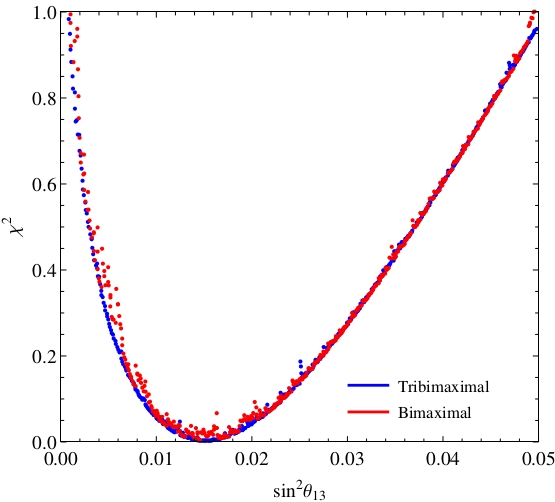}
\caption{\label{fig3}\it $\chi^2$ as a function of $\sin^2\theta_{13}$ in the  type-II see-saw $SO(10)$ models obtained 
  when starting in the TBM or BM basis.}  
\end{center}
\end{figure}

As we have already seen, 
in models of this type  neutrino masses are described by
${\cal M}_\nu~\sim~fv_L\;,$
where $v_L$ is the vev of the $B-L=2$ triplet in the ${\bf\overline{126}}_H$ Higgs
field and $f$ is its Yukawa coupling matrix with the $\bf 16$.
Since one can decide to work in a basis where the matrix $f$ is diagonalized by the BM or by TBM matrices, 
the results of a fit of the model parameters on the fermion observables performed in one basis lead to the same $\chi^2$ than the fit 
in the other basis, thus a  $\chi^2$ analysis cannot decide whether TBM or BM is a better starting point \cite{Altarelli:2015foa}. 
This is confirmed by the plot in Fig.\ref{fig3}, where it is shown that, within uncertainties, the $\chi^2$ as a function of the reactor angle is equal in the two cases, 
and this is true also for values of $\sin\theta_{13}$ different than the measured value. In particular, the minimum $\chi^2$ value, $\chi^2 =0.003$, is obtained for
  $\sin^2\theta_{13} \sim 0.015$, just a bit below the experimental value
  $\sin^2\theta_{13} \sim 0.022$.  Nevertheless, as the minimum
  $\chi^2$ is quite shallow for $\sin^2\theta_{13}<0.1$, the fit does not
  exhibit any strongly preferred value of $\theta_{13}$.

\begin{figure}[h!]
\begin{center}
\includegraphics[scale=.5]{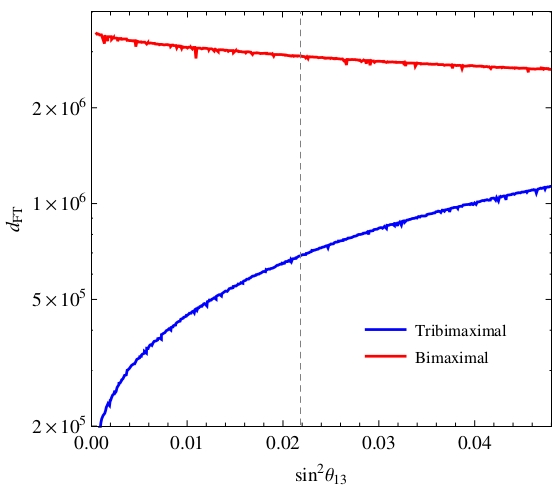}
\caption{\label{fig4}\it The behavior of the $d_{FT}$ increases (decreases) with $\sin^2{\theta_{13}}$ in the TBM (BM) cases. For the physical value $\sin^2{\theta_{13}} \sim 0.022$ it is about 4 times larger in the BM case.}
\end{center}
\end{figure}

Having established that the $\chi^2$ is not the best variable to decide whether TBM or BM is better, one can consider  
to measure the amount of fine-tuning needed to fit a set of data by means of the parameter $d_{FT}$  
introduced in Ref. \cite{Altarelli:2010at}:
\begin{equation}
\label{fine-tuning}
d_{FT} = \sum \left| \frac{p_i}{e_i} \right| \,,
\end{equation}
where $e_i$ is the "error" of a given parameter $p_i$ defined as the shift from the best fit value that changes 
the $\chi^2$ by one unit, with all other parameters fixed at their best 
fit values.
In Fig.\ref{fig4} we report a study of the fine tuning parameter   when the fit is repeated with the same data except for $\sin^2\theta_{13}$. 
It clearly shows that:
\begin{itemize}
 \item for the physical value of $\sin^2\theta_{13}$, $d_{FT}$ is smaller in the TBM case;
 \item the fine tuning increases (decreases) with $\sin\theta_{13}$ for TBM (BM).
\end{itemize}

A closer inspection of the $d_{FT}$ parameter reveals a series of interesting features: first of all, that
the large  values are predominantly driven by the smallness of the
electron mass; then,  due to the presence of mixing, the $d_{FT}$
coming from  the 33 component of $h$ (mainly
responsible for the top mass) is actually one of the largest
contributions to the global $d_{FT}$ because of its relevance to the
electron mass in both TBM and BM scenarios. 
Although
this might be surprising, one has to take into account that the
dependence of the observables on the parameters is quite complicated
due to the off-diagonal elements of the mass matrices.

Other classes of renormalizable and non-renormalizable $SO(10)$ models supplemented by discrete and continuous symmetries have been discussed in the literature. 
In \cite{Altarelli:2010at} a model comparison based on a $\chi^2$ analysis and on the values of  $d_{FT}$ has been carried out with sufficient details to allow for a discrimination in terms of performance in the description of the data. Tab.\ref{tab:compared} has been extracted from  \cite{Altarelli:2010at} and reports the results of such a comparison. The model called BSV \cite{Bajc:2001fe} (no flavor symmetries involved here) has a minimal Yukawa sector with ${\bf 10}_H$ and ${\bf\overline{126}}_H$ and has been compared with the data in Ref. \cite{Bertolini:2006pe}, where the type-I and mixed type-I and type-II cases were considered.  
As it is well known, the restricted Higgs content calls for  complex $h$ and $f$ matrices. Even increasing the number  of free parameters, with type-II dominance no good fit of the data can be obtained. 
The situation changes if one introduces the  ${\bf 120}_H$ of Higgs, as in the model with type-II see-saw dominance introduced by Joshipura and Kodrani (JK) \cite{Joshipura:2009tg}. The relevant feature of this model is the existence of a broken $\mu-\tau$ symmetry in addition to the   parity symmetry which causes hermitian mass matrices.  Similarly,  
Grimus and Kuhbock  \cite{Grimus:2006rk} (GK) also have an extended Higgs sector with ${\bf 10}_H$, ${\bf\overline{126}}$ and ${\bf 120}_H$ but their model is based on type-I see-saw dominance.

In the class of non-renormalizable $SO(10)$ theories, we can cite the model of Dermisek and Raby (DR)\cite{Dermisek:2005ij,Dermisek:2006dc}; it contains Higgses in the ${\bf 10}_H$, ${\bf 45}_H$ and ${\bf \overline{16}}_H$, and it is based on the flavour symmetry $S_3\times U(1) \times Z_2 \times Z_2$. 
In the symmetric $S_3$  limit  only the masses of the third generation are non-vanishing while the second and first generation masses are generated by a symmetry breaking stage. The neutrino masses are obtained through a type-I see-saw mechanism with a hierarchical Majorana mass matrix. Enough freedom to reproduce the observed neutrino properties is guarantee by new $SO(10)$-singlet neutrino and new scalar fields.

A similar Higgs sector with ${\bf 10}_H$, $({\bf 16 + \overline{16}})_H$ and ${\bf 45}_H$ representations and a few $SO(10)$ singlets constitute the scalar sector of the model by Albright, Babu and Barr  (ABB) \cite{Albright:2000dk,Albright:2001uh}. However, this model is based on a flavour symmetry $U(1)\times Z_2 \times Z_2$ which is mainly used to select the desired
terms which in the Lagrangian   and reject those that would not help in reproducing the data. 
A modification of this model has been  proposed by Ji, Li, Mohapatra (JLM) \cite{Ji:2005zk};  the charged lepton and the down quark  mass matrices are the same as in the ABB model but the up and Dirac neutrino mass matrices are modified thanks to new dimension five and six vertices introduced in the theory.  The model is based on type-I see-saw and the new operators provide a sufficient number of free parameters to fit the leptonic mixing angles.

\begin{table}[htbp]
\begin{center}
\begin{tabular}{|l|l|l|l|l|}
\hline
Model &d.o.f. &$\chi^2$ &  $\chi^2$/d.o.f. & $d_{FT}$  \\
\hline
\hline
DR    \cite{Dermisek:2005ij,Dermisek:2006dc} & 4 &0.41 & 0.10 &7.0 ~$10^3$\\
\hline
\hline
ABB    \cite{Albright:2000dk,Albright:2001uh} &6 & 2.8 & 0.47 &8.1~$10^3$\\
\hline
JLM    \cite{Ji:2005zk} &4 &2.9 & 0.74 & 9.4~$10^3$ \\
\hline
\hline
BSV    \cite{Bertolini:2006pe} &$< 0$ &6.9 &-& 2.0~$10^5$  \\
\hline
JK   \cite{Joshipura:2009tg} &3 &3.4 & 1.1 &4.7~$10^5$  \\
\hline
GK    \cite{Grimus:2006rk} &0 &0.15 & -& 1.5~$10^5$ \\
\hline
\end{tabular}
\end{center}
\caption{\it Comparison of different $SO(10)$ models fitted to the data. Above the double lines mark we report the non-renormalizable models whereas below we list the renormalizable models considered in this paper. Adapted from \cite{Altarelli:2010at}.}
\label{tab:compared}
\end{table}

The relevant feature of the results presented in  Tab.(\ref{tab:compared}) is that  the realistic $SO(10)$ models which are non renormalizable with type-I see-saw (DR, ABB, JLM), have a $\chi^2$/d.o.f. smaller than 1 and a moderate level of fine tuning  $d_{FT}$, if compared with the relatively more constrained BSV, JK and GK. They all have a large amount of fine tuning and, with the exception of the GK model, a worst $\chi^2$.
The larger fine tuning arises from the more pronounced difficulty of fitting the light first generation of charged fermion masses, together with the neutrino mass differences and mixing angles.

More recently, successful attempts to completely describe  neutrino data  within $S_4$ and $\Delta(27)$ have been presented in \cite{Bjorkeroth:2017ybg,Bjorkeroth:2015uou,Bjorkeroth:2016lzs}, where also the ability to provide a framework for the leptogenesis mechanism has been addressed \cite{Bjorkeroth:2016lzs}.

\begin{figure}[t]
\centering
\includegraphics[width=0.43\textwidth]{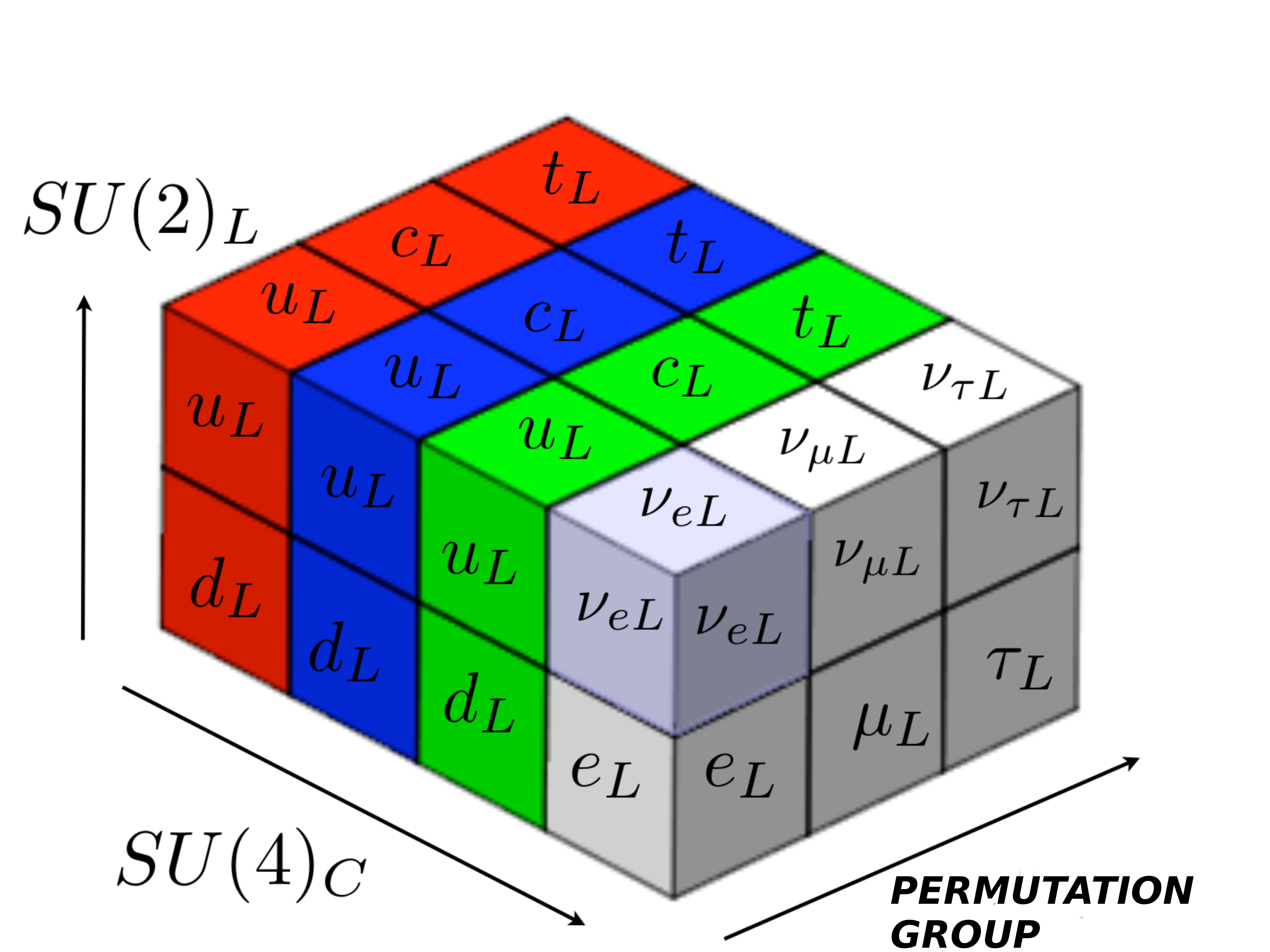}
\includegraphics[width=0.43\textwidth]{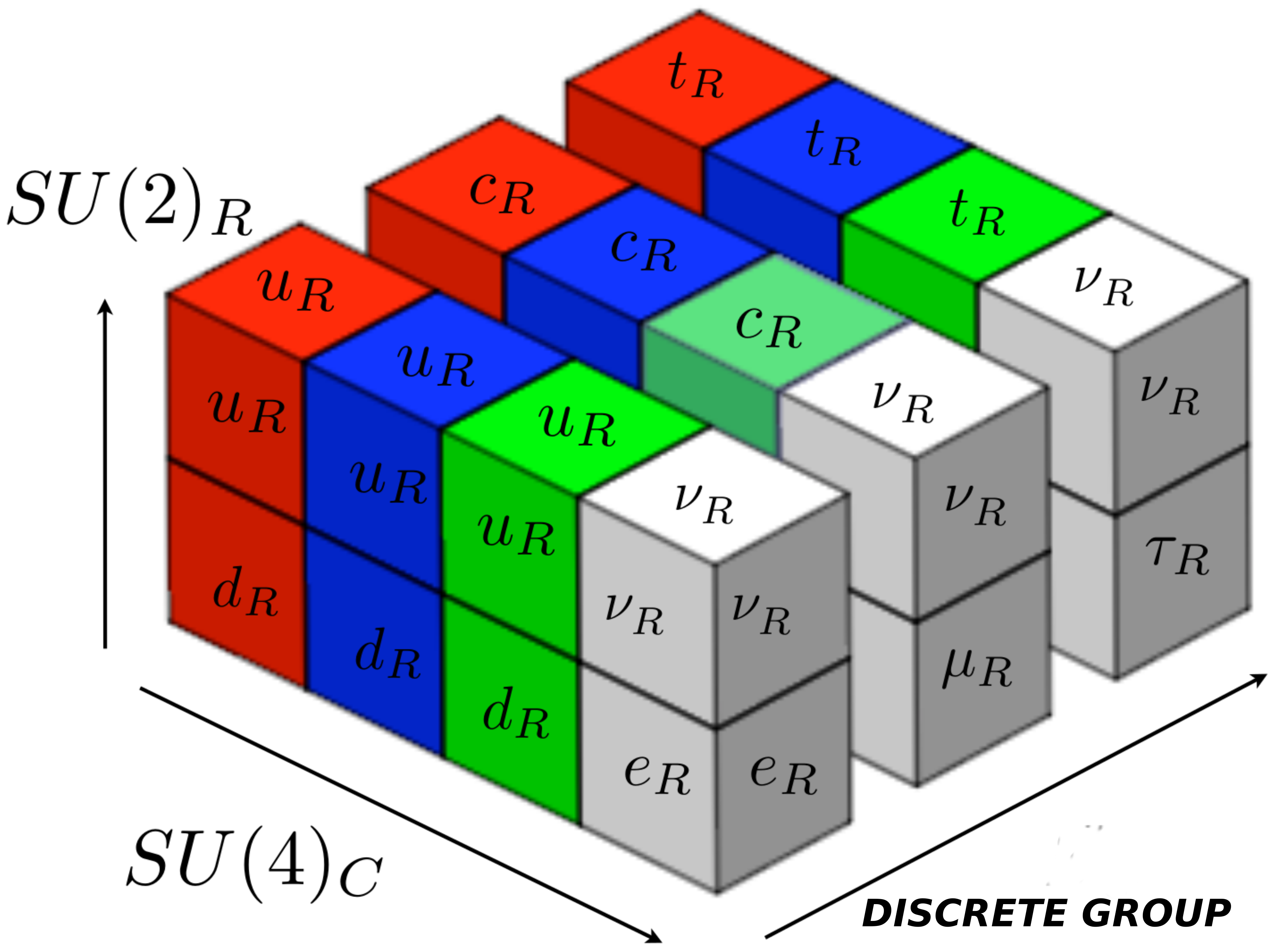}
    \caption{\it  \label{cubo} Pictorial representation of a possible particle assignment in models with $\left[ PS\otimes {\rm permutation} \otimes  {\rm discrete} \right]$ groups. Figure taken and modified from 
   \cite{King:2014iia}.}

\end{figure}

Beside the models with complete unification at the GUT scale, one can also consider the possibility of supplementing with flavor symmetries models with {\it partial} unification, that is theories where the gauge group at the GUT scale is not an unique group. Good examples in this direction are those based on the Pati-Salam group $SU(4)_c \otimes SU(2)_L \otimes SU(2)_R$ (PS), as discussed 
in \cite{Toorop:2010yh}, where $S_4$ was employed to recover the quark-lepton complementarity at LO and in \cite{King:2013hoa,King:2014iia}, which explores the capabilities of $A_4$ to describe 
quark and lepton masses, mixing and CP violation \footnote{See \cite{King:2006me} for an example of a  PS model where, instead of a discrete group, the continuos $SO(3)$ gauged family symmetry has been employed.}. As usual, these models also need the presence of additional discrete (or continuous $U(1)$) symmetries to forbid or suppress unwanted operators. 
In Fig.\ref{cubo}, modified from \cite{King:2014iia}, we sketch a possible particle assignment for models with $\left[ PS\otimes {\rm permutation} \otimes  {\rm discrete} \right]$ groups, where it is understood that the permutation group contains triplet representations.
In both panels, the red, blue and green colors represent the $SU(3)$ triplets, which are accompanied with the light gray particles to complete the fundamental ${\bf 4}$ representation of 
$SU(4)_c$.  The left-handed families are assigned to triplet presentations of the permutation groups and are doublets under $SU(2)_L$, left panel. On the right panel 
we consider that the right-handed families are distinguished by different charges  of the discrete group  and are doublets of $SU(2)_R$.

\section{Conclusions}
The question of the theoretical understanding of the experimental numbers of fermion masses and mixing is a very old story. Although neutrinos were considered as a promising tool to access the fundamental properties of particle interactions, the new data helped to discard some theoretical model on lepton mixing (mainly those based on $\theta_{13}=0$ at the LO) but many other still offer a viable solution,  spanning a wide range of possibilities going from a situation with no structure and no symmetry in the neutrino sector ({\it anarchy}) to a maximum of symmetry  for the models based on discrete non-abelian flavour groups. 

In this respect, neutrinos have not offered so far any crucial insight on the problem of flavour. The extension to include GUT (or Partial Unification) symmetry exacerbates the difficulties in the model building, as also the quark properties must be taken into account  and the larger symmetry reduces the useful number of free parameters.  

If one is driven by the fact that the quark-lepton complementarity is a real feature of Nature, then models based on $SU(5)$  with a broken $S_4$ symmetry emerge as one among the most viable and predictive theory, in which fermion masses and mixing are all well reproduced inside their experimental ranges at the prize of small fine-tunings in very few model parameters.

As we have seen in Tab.\ref{tab:oscillation_exp}, the octant of the atmospheric angle, the value of the CP violating phase $\delta$ and the neutrino mass orderings are features of the neutrinos that have not been clearly addressed so far. Thus, from the model building point of view, the results coming from the running (for instance, NO$\nu$A \cite{Adamson:2017gxd} and T2K \cite{Abe:2017uxa}) and planned experiments (like DUNE \cite{Acciarri:2016crz})
can certainly help in selecting the class of models that, more than others, will be able to incorporate the new information.   In this respect, the emerging indication of $\delta\sim 3/2 \pi$
seems to exclude the whole class of models predicting CP-conserving Dirac phase, as many do of those listed in Section \ref{CPsymm}.\\
On the other hand, the uncertainties affecting the already measured mixing angle and mass differences are expected to be reduced   to a sub-percent level in the next 5-10 years (as it is  the case for the solar parameters measured by the JUNO detector \cite{Djurcic:2015vqa}) and, in a framework where the mixing parameters are obtained from a LO neutrino mass texture corrected by charge lepton rotations, this can influence in a critical manner which LO  mass matrix is the most useful starting point; with more precise measurements,  the {\it jumps} described in Fig.\ref{f1}, needed to reconcile the LO predictions with the data, must be chosen more carefully.

\section{Acknowledgements}
The author is strongly indebted with Erica Vagnoni and Andrea Di Iura for useful discussions.

\bibliographystyle{MLA}

\begin{thebibliography}{1}


\bibitem{Majorana:1937vz}
  E.~Majorana,
  Nuovo Cim.\  {\bf 14} (1937) 171.
  doi:10.1007/BF02961314


\bibitem{Cabibbo:1963yz}
  N.~Cabibbo,
  Phys.\ Rev.\ Lett.\  {\bf 10} (1963) 531.
  doi:10.1103/PhysRevLett.10.531

  
\bibitem{Kobayashi:1973fv}
  M.~Kobayashi and T.~Maskawa,
  Prog.\ Theor.\ Phys.\  {\bf 49} (1973) 652.
  doi:10.1143/PTP.49.652
  
\bibitem{Pontecorvo:1957cp}
  B.~Pontecorvo,
  Sov.\ Phys.\ JETP {\bf 6} (1957) 429
   [Zh.\ Eksp.\ Teor.\ Fiz.\  {\bf 33} (1957) 549].

\bibitem{Pontecorvo:1957qd}
  B.~Pontecorvo,
  Sov.\ Phys.\ JETP {\bf 7} (1958) 172
   [Zh.\ Eksp.\ Teor.\ Fiz.\  {\bf 34} (1957) 247].

\bibitem{Maki:1962mu}
  Z.~Maki, M.~Nakagawa and S.~Sakata,
  Prog.\ Theor.\ Phys.\  {\bf 28} (1962) 870.
  doi:10.1143/PTP.28.870

\bibitem{Pontecorvo:1967fh}
  B.~Pontecorvo,
  Sov.\ Phys.\ JETP {\bf 26} (1968) 984
   [Zh.\ Eksp.\ Teor.\ Fiz.\  {\bf 53} (1967) 1717].

  
\bibitem{Raidal:2004iw}
  M.~Raidal,
  Phys.\ Rev.\ Lett.\  {\bf 93} (2004) 161801
  doi:10.1103/PhysRevLett.93.161801
  [hep-ph/0404046].
  
\bibitem{Antusch:2005ca}
  S.~Antusch, S.~F.~King and R.~N.~Mohapatra,
  Phys.\ Lett.\ B {\bf 618} (2005) 150
  doi:10.1016/j.physletb.2005.05.026
  [hep-ph/0504007].

\bibitem{Minakata:2004xt}
  H.~Minakata and A.~Y.~Smirnov,
  Phys.\ Rev.\ D {\bf 70} (2004) 073009
  doi:10.1103/PhysRevD.70.073009
  [hep-ph/0405088].

  
\bibitem{Frampton:2004vw}
  P.~H.~Frampton and R.~N.~Mohapatra,
  JHEP {\bf 0501} (2005) 025
  doi:10.1088/1126-6708/2005/01/025
  [hep-ph/0407139].
  
\bibitem{Harrison:2002er}
P.~F.~Harrison, D.~H.~Perkins and W.~G.~Scott,
Phys.\ Lett.\ B {\bf 530} (2002) 167
doi:10.1016/S0370-2693(02)01336-9
[hep-ph/0202074].

\bibitem{Harrison:2002kp}
P.~F.~Harrison and W.~G.~Scott,
Phys.\ Lett.\ B {\bf 535} (2002) 163
doi:10.1016/S0370-2693(02)01753-7
[hep-ph/0203209].

\bibitem{Xing:2002sw}
Z.~z.~Xing,
Phys.\ Lett.\ B {\bf 533} (2002) 85
doi:10.1016/S0370-2693(02)01649-0
[hep-ph/0204049].

\bibitem{Harrison:2002et}
P.~F.~Harrison and W.~G.~Scott,
Phys.\ Lett.\ B {\bf 547} (2002) 219
doi:10.1016/S0370-2693(02)02772-7
[hep-ph/0210197].
  
\bibitem{Harrison:2003aw}
  P.~F.~Harrison and W.~G.~Scott,
  Phys.\ Lett.\ B {\bf 557} (2003) 76
  doi:10.1016/S0370-2693(03)00183-7
  [hep-ph/0302025].
  
  
\bibitem{Froggatt:1978nt}
C.~D.~Froggatt and H.~B.~Nielsen,
Nucl.\ Phys.\ B {\bf 147} (1979) 277.
doi:10.1016/0550-3213(79)90316-X
  
\bibitem{Luhn:2011ip}
C.~Luhn,
JHEP {\bf 1103} (2011) 108
doi:10.1007/JHEP03(2011)108
[arXiv:1101.2417 [hep-ph]].
  
\bibitem{Altarelli:2006kg}
G.~Altarelli, F.~Feruglio and Y.~Lin,
Nucl.\ Phys.\ B {\bf 775} (2007) 31
doi:10.1016/j.nuclphysb.2007.03.042
[hep-ph/0610165].

  
 
\bibitem{Esteban:2016qun} 
  I.~Esteban, M.~C.~Gonzalez-Garcia, M.~Maltoni, I.~Martinez-Soler and T.~Schwetz,
  JHEP {\bf 1701}, 087 (2017)
  doi:10.1007/JHEP01(2017)087
  [arXiv:1611.01514 [hep-ph]].
  
  

\bibitem{Ibarra:2016qnf} 
  A.~Ibarra,
  Acta Phys.\ Polon.\ Supp.\  {\bf 9}, 741 (2016).
  doi:10.5506/APhysPolBSupp.9.741
\bibitem{Minkowski:1977sc} 
  P.~Minkowski,
  Phys.\ Lett.\  {\bf 67B}, 421 (1977).
  doi:10.1016/0370-2693(77)90435-X
\bibitem{Yanagida:1979}
Yanagida T 1979  In Proceedings of the Workshop on the Baryon Number of the
  Universe and Unified Theories, Tsukuba, Japan, 13-14 Feb 1979
\bibitem{Gell-Mann:1979}
Gell-Mann M, Ramond P and Slansky R 1979  Print-80-0576 (CERN)

\bibitem{Glashow:1980}
Glashow S~L 1980  In Quarks and Leptons, Cargese, ed. M. Levy et al., Plenum,
  1980 New York, p. 707

\bibitem{Mohapatra:1980}
Mohapatra R~N and Senjanovic G 1980 {\em Phys. Rev. Lett.\/} {\bf 44} 912


\bibitem{Konetschny:1977bn} 
  W.~Konetschny and W.~Kummer,
  Phys.\ Lett.\  {\bf 70B}, 433 (1977).
  doi:10.1016/0370-2693(77)90407-5
\bibitem{Foot:1988aq} 
  R.~Foot, H.~Lew, X.~G.~He and G.~C.~Joshi,
  Z.\ Phys.\ C {\bf 44}, 441 (1989).
  doi:10.1007/BF01415558
\bibitem{Altarelli:2004za} 
  G.~Altarelli and F.~Feruglio,
  New J.\ Phys.\  {\bf 6}, 106 (2004)
  doi:10.1088/1367-2630/6/1/106
  [hep-ph/0405048].

\bibitem{Fritzsch:1974nn}
  H.~Fritzsch and P.~Minkowski,
  Annals Phys.\  {\bf 93} (1975) 193.
  doi:10.1016/0003-4916(75)90211-0

  
\bibitem{delAguila:1980qag}
  F.~del Aguila and L.~E.~Ibanez,
  Nucl.\ Phys.\ B {\bf 177} (1981) 60.
  doi:10.1016/0550-3213(81)90266-2

\bibitem{Deshpande:1992au}
  N.~G.~Deshpande, E.~Keith and P.~B.~Pal,
  Phys.\ Rev.\ D {\bf 46} (1993) 2261.
  doi:10.1103/PhysRevD.46.2261
  

\bibitem{Pati:1974yy}
  J.~C.~Pati and A.~Salam,
  Phys.\ Rev.\ D {\bf 10} (1974) 275
   Erratum: [Phys.\ Rev.\ D {\bf 11} (1975) 703].
  doi:10.1103/PhysRevD.10.275, 10.1103/PhysRevD.11.703.2
  

\bibitem{Georgi:1974sy}
  H.~Georgi and S.~L.~Glashow,
  Phys.\ Rev.\ Lett.\  {\bf 32} (1974) 438.
  doi:10.1103/PhysRevLett.32.438
  
  
  
  
\bibitem{Altarelli:2013aqa}
G.~Altarelli and D.~Meloni,
JHEP {\bf 1308} (2013) 021
doi:10.1007/JHEP08(2013)021
[arXiv:1305.1001 [hep-ph]].


\bibitem{Deppisch:2017xhv}
  F.~F.~Deppisch, T.~E.~Gonzalo and L.~Graf,
  arXiv:1705.05416 [hep-ph].


\bibitem{Bajc:2005zf}
  B.~Bajc, A.~Melfo, G.~Senjanovic and F.~Vissani,
  Phys.\ Rev.\ D {\bf 73} (2006) 055001
  doi:10.1103/PhysRevD.73.055001
  [hep-ph/0510139].
  
\bibitem{Peccei:1977hh}
  R.~D.~Peccei and H.~R.~Quinn,
  Phys.\ Rev.\ Lett.\  {\bf 38} (1977) 1440.
  doi:10.1103/PhysRevLett.38.1440
  
\bibitem{Babu:1992ia}
  K.~S.~Babu and R.~N.~Mohapatra,
  Phys.\ Rev.\ Lett.\  {\bf 70} (1993) 2845
  doi:10.1103/PhysRevLett.70.2845
  [hep-ph/9209215].
  
  
  
  
  


  

  

  





  

  
\bibitem{Georgi:1979df}
  H.~Georgi and C.~Jarlskog,
  Phys.\ Lett.\  {\bf 86B} (1979) 297.
  doi:10.1016/0370-2693(79)90842-6

\bibitem{Joshipura:2011nn}
A.~S.~Joshipura and K.~M.~Patel,
Phys.\ Rev.\ D {\bf 83} (2011) 095002
doi:10.1103/PhysRevD.83.095002
[arXiv:1102.5148 [hep-ph]].


\bibitem{Dueck:2013gca}
A.~Dueck and W.~Rodejohann,
JHEP {\bf 1309} (2013) 024
doi:10.1007/JHEP09(2013)024
[arXiv:1306.4468 [hep-ph]].

\bibitem{Arason:1992eb}
  H.~Arason, D.~J.~Castano, E.~J.~Piard and P.~Ramond,
  Phys.\ Rev.\ D {\bf 47} (1993) 232
  doi:10.1103/PhysRevD.47.232
  [hep-ph/9204225].

\bibitem{Harvey:1980je}
J.~A.~Harvey, P.~Ramond and D.~B.~Reiss,
Phys.\ Lett.\  {\bf 92B} (1980) 309.
doi:10.1016/0370-2693(80)90270-1

  
  
\bibitem{Harvey:1981hk}
  J.~A.~Harvey, D.~B.~Reiss and P.~Ramond,
  Nucl.\ Phys.\ B {\bf 199} (1982) 223.
  doi:10.1016/0550-3213(82)90346-7
  
\bibitem{Matsuda:1999yx}
  K.~Matsuda, T.~Fukuyama and H.~Nishiura,
  Phys.\ Rev.\ D {\bf 61} (2000) 053001
  doi:10.1103/PhysRevD.61.053001
  [hep-ph/9906433].
 

   
\bibitem{Bajc:2001fe}
  B.~Bajc, G.~Senjanovic and F.~Vissani,
  PoS HEP {\bf 2001} (2001) 198
  [hep-ph/0110310].

\bibitem{Bajc:2002iw}
  B.~Bajc, G.~Senjanovic and F.~Vissani,
  Phys.\ Rev.\ Lett.\  {\bf 90} (2003) 051802
  doi:10.1103/PhysRevLett.90.051802
  [hep-ph/0210207].

\bibitem{Meloni:2014rga}
D.~Meloni, T.~Ohlsson and S.~Riad,
JHEP {\bf 1412} (2014) 052
doi:10.1007/JHEP12(2014)052
[arXiv:1409.3730 [hep-ph]].

\bibitem{Meloni:2016rnt}
D.~Meloni, T.~Ohlsson and S.~Riad,
JHEP {\bf 1703} (2017) 045
doi:10.1007/JHEP03(2017)045
[arXiv:1612.07973 [hep-ph]].

\bibitem{Xing:2007fb}
  Z.~z.~Xing, H.~Zhang and S.~Zhou,
  Phys.\ Rev.\ D {\bf 77} (2008) 113016
  doi:10.1103/PhysRevD.77.113016
  [arXiv:0712.1419 [hep-ph]].
  
  
\bibitem{Bertolini:2006pe}
  S.~Bertolini, T.~Schwetz and M.~Malinsky,
  Phys.\ Rev.\ D {\bf 73} (2006) 115012
  doi:10.1103/PhysRevD.73.115012
  [hep-ph/0605006].
  
\bibitem{Lavoura:2006dv}
  L.~Lavoura, H.~Kuhbock and W.~Grimus,
  Nucl.\ Phys.\ B {\bf 754} (2006) 1
  doi:10.1016/j.nuclphysb.2006.07.024
  [hep-ph/0603259].

\bibitem{Bajc:2005aq}
  B.~Bajc and G.~Senjanovic,
  Phys.\ Rev.\ Lett.\  {\bf 95} (2005) 261804
  doi:10.1103/PhysRevLett.95.261804
  [hep-ph/0507169].

  
\bibitem{Lam:2007qc}
C.~S.~Lam,
Phys.\ Lett.\ B {\bf 656} (2007) 193
doi:10.1016/j.physletb.2007.09.032
[arXiv:0708.3665 [hep-ph]].
  
\bibitem{Lam:2008rs}
C.~S.~Lam,
Phys.\ Rev.\ Lett.\  {\bf 101} (2008) 121602
doi:10.1103/PhysRevLett.101.121602
[arXiv:0804.2622 [hep-ph]].
  
\bibitem{Lam:2008sh}
C.~S.~Lam,
Phys.\ Rev.\ D {\bf 78} (2008) 073015
doi:10.1103/PhysRevD.78.073015
[arXiv:0809.1185 [hep-ph]].
  
\bibitem{Fonseca:2014koa}
  R.~M.~Fonseca and W.~Grimus,
  JHEP {\bf 1409} (2014) 033
  doi:10.1007/JHEP09(2014)033
  [arXiv:1405.3678 [hep-ph]].
  

  
\bibitem{Hernandez:2012ra}
  D.~Hernandez and A.~Y.~Smirnov,
  Phys.\ Rev.\ D {\bf 86} (2012) 053014
  doi:10.1103/PhysRevD.86.053014
  [arXiv:1204.0445 [hep-ph]].
  
\bibitem{Hernandez:2012sk}
  D.~Hernandez and A.~Y.~Smirnov,
  Phys.\ Rev.\ D {\bf 87} (2013) no.5,  053005
  doi:10.1103/PhysRevD.87.053005
  [arXiv:1212.2149 [hep-ph]].

\bibitem{Grimus:2013rw}
  W.~Grimus,
  J.\ Phys.\ G {\bf 40} (2013) 075008
  doi:10.1088/0954-3899/40/7/075008
  [arXiv:1301.0495 [hep-ph]].
  
\bibitem{Ge:2011ih}
  S.~F.~Ge, D.~A.~Dicus and W.~W.~Repko,
  Phys.\ Lett.\ B {\bf 702} (2011) 220
  doi:10.1016/j.physletb.2011.06.096
  [arXiv:1104.0602 [hep-ph]].
  
\bibitem{Ge:2011qn}
  S.~F.~Ge, D.~A.~Dicus and W.~W.~Repko,
  Phys.\ Rev.\ Lett.\  {\bf 108} (2012) 041801
  doi:10.1103/PhysRevLett.108.041801
  [arXiv:1108.0964 [hep-ph]].
  
  
\bibitem{King:2001uz}
  S.~F.~King and G.~G.~Ross,
  Phys.\ Lett.\ B {\bf 520} (2001) 243
  doi:10.1016/S0370-2693(01)01139-X
  [hep-ph/0108112].
  
\bibitem{King:2013eh}
  S.~F.~King and C.~Luhn,
  Rept.\ Prog.\ Phys.\  {\bf 76} (2013) 056201
  doi:10.1088/0034-4885/76/5/056201
  [arXiv:1301.1340 [hep-ph]].

  
\bibitem{Barry:2010yk}
  J.~Barry and W.~Rodejohann,
  Nucl.\ Phys.\ B {\bf 842} (2011) 33
  doi:10.1016/j.nuclphysb.2010.08.015
  [arXiv:1007.5217 [hep-ph]].
  
  
  
\bibitem{Ballett:2013wya}
  P.~Ballett, S.~F.~King, C.~Luhn, S.~Pascoli and M.~A.~Schmidt,
  Phys.\ Rev.\ D {\bf 89} (2014) no.1,  016016
  doi:10.1103/PhysRevD.89.016016
  [arXiv:1308.4314 [hep-ph]].

  
  
\bibitem{Meloni:2013qda}
  D.~Meloni,
  Phys.\ Lett.\ B {\bf 728} (2014) 118
  doi:10.1016/j.physletb.2013.11.033
  [arXiv:1308.4578 [hep-ph]].

  
\bibitem{Petcov:2014laa}
  S.~T.~Petcov,
  Nucl.\ Phys.\ B {\bf 892} (2015) 400
  doi:10.1016/j.nuclphysb.2015.01.011
  [arXiv:1405.6006 [hep-ph]].

  
\bibitem{Girardi:2014faa}
  I.~Girardi, S.~T.~Petcov and A.~V.~Titov,
  Nucl.\ Phys.\ B {\bf 894} (2015) 733
  doi:10.1016/j.nuclphysb.2015.03.026
  [arXiv:1410.8056 [hep-ph]].
  
\bibitem{Girardi:2015vha}
  I.~Girardi, S.~T.~Petcov and A.~V.~Titov,
  Eur.\ Phys.\ J.\ C {\bf 75} (2015) 345
  doi:10.1140/epjc/s10052-015-3559-6
  [arXiv:1504.00658 [hep-ph]].
  
\bibitem{Girardi:2015rwa}
  I.~Girardi, S.~T.~Petcov, A.~J.~Stuart and A.~V.~Titov,
  Nucl.\ Phys.\ B {\bf 902} (2016) 1
  doi:10.1016/j.nuclphysb.2015.10.020
  [arXiv:1509.02502 [hep-ph]].

  
  
\bibitem{Altarelli:2005yx}
  G.~Altarelli and F.~Feruglio,
  Nucl.\ Phys.\ B {\bf 741} (2006) 215
  doi:10.1016/j.nuclphysb.2006.02.015
  [hep-ph/0512103].
  
  
\bibitem{deMedeirosVarzielas:2005qg}
  I.~de Medeiros Varzielas, S.~F.~King and G.~G.~Ross,
  Phys.\ Lett.\ B {\bf 644} (2007) 153
  doi:10.1016/j.physletb.2006.11.015
  [hep-ph/0512313].
 
\bibitem{Grimus:2005mu}
  W.~Grimus and L.~Lavoura,
  JHEP {\bf 0508} (2005) 013
  doi:10.1088/1126-6708/2005/08/013
  [hep-ph/0504153].

\bibitem{Meloni:2010aw}
  D.~Meloni, S.~Morisi and E.~Peinado,
  J.\ Phys.\ G {\bf 38} (2011) 015003
  doi:10.1088/0954-3899/38/1/015003
  [arXiv:1005.3482 [hep-ph]].

  
\bibitem{Ferreira:2012ri}
  P.~M.~Ferreira, W.~Grimus, L.~Lavoura and P.~O.~Ludl,
  JHEP {\bf 1209} (2012) 128
  doi:10.1007/JHEP09(2012)128
  [arXiv:1206.7072 [hep-ph]].
 
 
\bibitem{Kobayashi:2008ih}
  T.~Kobayashi, Y.~Omura and K.~Yoshioka,
  Phys.\ Rev.\ D {\bf 78} (2008) 115006
  doi:10.1103/PhysRevD.78.115006
  [arXiv:0809.3064 [hep-ph]].

\bibitem{Burrows:2010wz}
  T.~J.~Burrows and S.~F.~King,
  Nucl.\ Phys.\ B {\bf 842} (2011) 107
  doi:10.1016/j.nuclphysb.2010.08.018
  [arXiv:1007.2310 [hep-ph]].

  

\bibitem{Altarelli:2010gt}
  G.~Altarelli and F.~Feruglio,
  Rev.\ Mod.\ Phys.\  {\bf 82} (2010) 2701
  doi:10.1103/RevModPhys.82.2701
  [arXiv:1002.0211 [hep-ph]].
  
  
\bibitem{Fukuyama:1997ky}
  T.~Fukuyama and H.~Nishiura,
  hep-ph/9702253.
  
\bibitem{Fukuyama:2017qxb}
  T.~Fukuyama,
  PTEP {\bf 2017} (2017) no.3,  033B11
  doi:10.1093/ptep/ptx032
  [arXiv:1701.04985 [hep-ph]].

  
  
\bibitem{Barger:1998ta}
V.~D.~Barger, S.~Pakvasa, T.~J.~Weiler and K.~Whisnant,
Phys.\ Lett.\ B {\bf 437} (1998) 107
doi:10.1016/S0370-2693(98)00880-6
[hep-ph/9806387].
  
  


  
  
  
\bibitem{Datta:2003qg}
A.~Datta, F.~S.~Ling and P.~Ramond,
Nucl.\ Phys.\ B {\bf 671} (2003) 383
doi:10.1016/j.nuclphysb.2003.08.026
[hep-ph/0306002].

\bibitem{Kajiyama:2007gx}
Y.~Kajiyama, M.~Raidal and A.~Strumia,
Phys.\ Rev.\ D {\bf 76} (2007) 117301
doi:10.1103/PhysRevD.76.117301
[arXiv:0705.4559 [hep-ph]].
  
\bibitem{Everett:2008et}
L.~L.~Everett and A.~J.~Stuart,
Phys.\ Rev.\ D {\bf 79} (2009) 085005
doi:10.1103/PhysRevD.79.085005
[arXiv:0812.1057 [hep-ph]].

\bibitem{Feruglio:2011qq}
F.~Feruglio and A.~Paris,
JHEP {\bf 1103} (2011) 101
doi:10.1007/JHEP03(2011)101
[arXiv:1101.0393 [hep-ph]].


\bibitem{Rodejohann:2008ir}
  W.~Rodejohann,
  Phys.\ Lett.\ B {\bf 671} (2009) 267
  doi:10.1016/j.physletb.2008.12.010
  [arXiv:0810.5239 [hep-ph]].
  
\bibitem{Capozzi:2016rtj}
F.~Capozzi, E.~Lisi, A.~Marrone, D.~Montanino and A.~Palazzo,
Nucl.\ Phys.\ B {\bf 908} (2016) 218
doi:10.1016/j.nuclphysb.2016.02.016
[arXiv:1601.07777 [hep-ph]].

\bibitem{Forero:2014bxa}
D.~V.~Forero, M.~Tortola and J.~W.~F.~Valle,
Phys.\ Rev.\ D {\bf 90} (2014) no.9,  093006
doi:10.1103/PhysRevD.90.093006
[arXiv:1405.7540 [hep-ph]].

\bibitem{Gonzalez-Garcia:2014bfa}
M.~C.~Gonzalez-Garcia, M.~Maltoni and T.~Schwetz,
JHEP {\bf 1411} (2014) 052
doi:10.1007/JHEP11(2014)052
[arXiv:1409.5439 [hep-ph]].




  
  
  
  
  
  
  
  
  
  
  
  
  
  
  

\bibitem{Petcov:2004rk}
  S.~T.~Petcov and W.~Rodejohann,
  Phys.\ Rev.\ D {\bf 71} (2005) 073002
  doi:10.1103/PhysRevD.71.073002
  [hep-ph/0409135].
  
\bibitem{Altarelli:2004jb}
  G.~Altarelli, F.~Feruglio and I.~Masina,
  Nucl.\ Phys.\ B {\bf 689} (2004) 157
  doi:10.1016/j.nuclphysb.2004.04.012
  [hep-ph/0402155].

\bibitem{Meloni:2010cj}
  D.~Meloni, F.~Plentinger and W.~Winter,
  Phys.\ Lett.\ B {\bf 699} (2011) 354
  doi:10.1016/j.physletb.2011.04.033
  [arXiv:1012.1618 [hep-ph]].
  
  
 
 
\bibitem{Ma:2005qf}
  E.~Ma,
  Phys.\ Rev.\ D {\bf 73} (2006) 057304
  doi:10.1103/PhysRevD.73.057304
  [hep-ph/0511133].

\bibitem{He:2006dk}
  X.~G.~He, Y.~Y.~Keum and R.~R.~Volkas,
  JHEP {\bf 0604} (2006) 039
  doi:10.1088/1126-6708/2006/04/039
  [hep-ph/0601001].
  
\bibitem{Chen:2009um}
  M.~C.~Chen and S.~F.~King,
  JHEP {\bf 0906} (2009) 072
  doi:10.1088/1126-6708/2009/06/072
  [arXiv:0903.0125 [hep-ph]].
  
  
\bibitem{Altarelli:2009kr}
  G.~Altarelli and D.~Meloni,
  J.\ Phys.\ G {\bf 36} (2009) 085005
  doi:10.1088/0954-3899/36/8/085005
  [arXiv:0905.0620 [hep-ph]].

 
\bibitem{Bazzocchi:2009pv}
  F.~Bazzocchi, L.~Merlo and S.~Morisi,
  Nucl.\ Phys.\ B {\bf 816} (2009) 204
  doi:10.1016/j.nuclphysb.2009.03.005
  [arXiv:0901.2086 [hep-ph]].

\bibitem{Grimus:2009pg}
  W.~Grimus, L.~Lavoura and P.~O.~Ludl,
  J.\ Phys.\ G {\bf 36} (2009) 115007
  doi:10.1088/0954-3899/36/11/115007
  [arXiv:0906.2689 [hep-ph]].
  
  
\bibitem{Aranda:2007dp}
  A.~Aranda,
  Phys.\ Rev.\ D {\bf 76} (2007) 111301
  doi:10.1103/PhysRevD.76.111301
  [arXiv:0707.3661 [hep-ph]].
 
\bibitem{Ding:2008rj}
  G.~J.~Ding,
  Phys.\ Rev.\ D {\bf 78} (2008) 036011
  doi:10.1103/PhysRevD.78.036011
  [arXiv:0803.2278 [hep-ph]].

\bibitem{Frampton:2008bz}
  P.~H.~Frampton, T.~W.~Kephart and S.~Matsuzaki,
  Phys.\ Rev.\ D {\bf 78} (2008) 073004
  doi:10.1103/PhysRevD.78.073004
  [arXiv:0807.4713 [hep-ph]].
  
 
\bibitem{Mohapatra:1998ka}
  R.~N.~Mohapatra and S.~Nussinov,
  Phys.\ Rev.\ D {\bf 60} (1999) 013002
  doi:10.1103/PhysRevD.60.013002
  [hep-ph/9809415].
  
\bibitem{Altarelli:2009gn}
  G.~Altarelli, F.~Feruglio and L.~Merlo,
  JHEP {\bf 0905} (2009) 020
  doi:10.1088/1126-6708/2009/05/020
  [arXiv:0903.1940 [hep-ph]].

\bibitem{Adulpravitchai:2009bg}
  A.~Adulpravitchai, A.~Blum and W.~Rodejohann,
  New J.\ Phys.\  {\bf 11} (2009) 063026
  doi:10.1088/1367-2630/11/6/063026
  [arXiv:0903.0531 [hep-ph]].
 

\bibitem{Albright:2010ap}
C.~H.~Albright, A.~Dueck and W.~Rodejohann,
Eur.\ Phys.\ J.\ C {\bf 70} (2010) 1099
doi:10.1140/epjc/s10052-010-1492-2
[arXiv:1004.2798 [hep-ph]].

\bibitem{Kim:2010zub}
J.~E.~Kim and M.~S.~Seo,
JHEP {\bf 1102} (2011) 097
doi:10.1007/JHEP02(2011)097
[arXiv:1005.4684 [hep-ph]].

\bibitem{Ishimori:2010au}
  H.~Ishimori, T.~Kobayashi, H.~Ohki, Y.~Shimizu, H.~Okada and M.~Tanimoto,
  Prog.\ Theor.\ Phys.\ Suppl.\  {\bf 183} (2010) 1
  doi:10.1143/PTPS.183.1
  [arXiv:1003.3552 [hep-th]].

  
\bibitem{Grimus:2011fk}
  W.~Grimus and P.~O.~Ludl,
  J.\ Phys.\ A {\bf 45} (2012) 233001
  doi:10.1088/1751-8113/45/23/233001
  [arXiv:1110.6376 [hep-ph]].
  
\bibitem{deAdelhartToorop:2011re}
R.~de Adelhart Toorop, F.~Feruglio and C.~Hagedorn,
Nucl.\ Phys.\ B {\bf 858} (2012) 437
doi:10.1016/j.nuclphysb.2012.01.017
[arXiv:1112.1340 [hep-ph]].
  
  
\bibitem{Jarlskog:1985ht}
  C.~Jarlskog,
  Phys.\ Rev.\ Lett.\  {\bf 55} (1985) 1039.
  doi:10.1103/PhysRevLett.55.1039

  
  



\bibitem{Lam:2011ag}
  C.~S.~Lam,
  Phys.\ Rev.\ D {\bf 83} (2011) 113002
  doi:10.1103/PhysRevD.83.113002
  [arXiv:1104.0055 [hep-ph]].


  
  
\bibitem{Grimus:2008tt}
  W.~Grimus and L.~Lavoura,
  JHEP {\bf 0809} (2008) 106
  doi:10.1088/1126-6708/2008/09/106
  [arXiv:0809.0226 [hep-ph]].
  
  
\bibitem{He:2006qd}
  X.~G.~He and A.~Zee,
  Phys.\ Lett.\ B {\bf 645} (2007) 427
  doi:10.1016/j.physletb.2006.11.055
  [hep-ph/0607163].
 

  
\bibitem{Antusch:2011ic}
  S.~Antusch, S.~F.~King, C.~Luhn and M.~Spinrath,
  Nucl.\ Phys.\ B {\bf 856} (2012) 328
  doi:10.1016/j.nuclphysb.2011.11.009
  [arXiv:1108.4278 [hep-ph]].
  
  
\bibitem{Bazzocchi:2011ax}
  F.~Bazzocchi,
  arXiv:1108.2497 [hep-ph].

  
\bibitem{King:2012in}
  S.~F.~King, C.~Luhn and A.~J.~Stuart,
  Nucl.\ Phys.\ B {\bf 867} (2013) 203
  doi:10.1016/j.nuclphysb.2012.09.021
  [arXiv:1207.5741 [hep-ph]].
  
  
\bibitem{Feruglio:2012cw}
  F.~Feruglio, C.~Hagedorn and R.~Ziegler,
  JHEP {\bf 1307} (2013) 027
  doi:10.1007/JHEP07(2013)027
  [arXiv:1211.5560 [hep-ph]].
  
\bibitem{Holthausen:2012dk}
  M.~Holthausen, M.~Lindner and M.~A.~Schmidt,
  JHEP {\bf 1304} (2013) 122
  doi:10.1007/JHEP04(2013)122
  [arXiv:1211.6953 [hep-ph]].

\bibitem{Chen:2014tpa}
  M.~C.~Chen, M.~Fallbacher, K.~T.~Mahanthappa, M.~Ratz and A.~Trautner,
  Nucl.\ Phys.\ B {\bf 883} (2014) 267
  doi:10.1016/j.nuclphysb.2014.03.023
  [arXiv:1402.0507 [hep-ph]].
  
  
\bibitem{Ecker:1983hz}
  G.~Ecker, W.~Grimus and H.~Neufeld,
  Nucl.\ Phys.\ B {\bf 247} (1984) 70.
  doi:10.1016/0550-3213(84)90373-0
  
\bibitem{Ecker:1987qp}
  G.~Ecker, W.~Grimus and H.~Neufeld,
  J.\ Phys.\ A {\bf 20} (1987) L807.
  doi:10.1088/0305-4470/20/12/010
  
  
\bibitem{Neufeld:1987wa}
  H.~Neufeld, W.~Grimus and G.~Ecker,
  Int.\ J.\ Mod.\ Phys.\ A {\bf 3} (1988) 603.
  doi:10.1142/S0217751X88000254
  
\bibitem{DiIura:2015kfa}
  A.~Di Iura, C.~Hagedorn and D.~Meloni,
  JHEP {\bf 1508} (2015) 037
  doi:10.1007/JHEP08(2015)037
  [arXiv:1503.04140 [hep-ph]].
  
\bibitem{Mohapatra:2012tb}
  R.~N.~Mohapatra and C.~C.~Nishi,
  Phys.\ Rev.\ D {\bf 86} (2012) 073007
  doi:10.1103/PhysRevD.86.073007
  [arXiv:1208.2875 [hep-ph]].
  
  
\bibitem{Feruglio:2013hia}
  F.~Feruglio, C.~Hagedorn and R.~Ziegler,
  Eur.\ Phys.\ J.\ C {\bf 74} (2014) 2753
  doi:10.1140/epjc/s10052-014-2753-2
  [arXiv:1303.7178 [hep-ph]].

\bibitem{Luhn:2013vna}
  C.~Luhn,
  Nucl.\ Phys.\ B {\bf 875} (2013) 80
  doi:10.1016/j.nuclphysb.2013.07.003
  [arXiv:1306.2358 [hep-ph]].
 
\bibitem{Penedo:2017vtf}
  J.~T.~Penedo, S.~T.~Petcov and A.~V.~Titov,
  arXiv:1705.00309 [hep-ph].
 


\bibitem{Li:2015jxa}
  C.~C.~Li and G.~J.~Ding,
  JHEP {\bf 1505} (2015) 100
  doi:10.1007/JHEP05(2015)100
  [arXiv:1503.03711 [hep-ph]].
  
  
\bibitem{Ballett:2015wia}
  P.~Ballett, S.~Pascoli and J.~Turner,
  Phys.\ Rev.\ D {\bf 92} (2015) no.9,  093008
  doi:10.1103/PhysRevD.92.093008
  [arXiv:1503.07543 [hep-ph]].

\bibitem{Turner:2015uta}
  J.~Turner,
  Phys.\ Rev.\ D {\bf 92} (2015) no.11,  116007
  doi:10.1103/PhysRevD.92.116007
  [arXiv:1507.06224 [hep-ph]].
  
\bibitem{deMedeirosVarzielas:2011zw}
  I.~de Medeiros Varzielas and D.~Emmanuel-Costa,
  Phys.\ Rev.\ D {\bf 84} (2011) 117901
  doi:10.1103/PhysRevD.84.117901
  [arXiv:1106.5477 [hep-ph]].
  
  
\bibitem{Bhattacharyya:2012pi}
  G.~Bhattacharyya, I.~de Medeiros Varzielas and P.~Leser,
  Phys.\ Rev.\ Lett.\  {\bf 109} (2012) 241603
  doi:10.1103/PhysRevLett.109.241603
  [arXiv:1210.0545 [hep-ph]].
 
\bibitem{Ma:2013xqa}
  E.~Ma,
  Phys.\ Lett.\ B {\bf 723} (2013) 161
  doi:10.1016/j.physletb.2013.05.011
  [arXiv:1304.1603 [hep-ph]].

\bibitem{Hagedorn:2014wha}
  C.~Hagedorn, A.~Meroni and E.~Molinaro,
  Nucl.\ Phys.\ B {\bf 891} (2015) 499
  doi:10.1016/j.nuclphysb.2014.12.013
  [arXiv:1408.7118 [hep-ph]].
  
\bibitem{Ding:2015rwa}
  G.~J.~Ding and S.~F.~King,
  Phys.\ Rev.\ D {\bf 93} (2016) 025013
  doi:10.1103/PhysRevD.93.025013
  [arXiv:1510.03188 [hep-ph]].
  

\bibitem{Altarelli:2012ia}
  G.~Altarelli, F.~Feruglio, I.~Masina and L.~Merlo,
  JHEP {\bf 1211} (2012) 139
  doi:10.1007/JHEP11(2012)139
  [arXiv:1207.0587 [hep-ph]].
  
\bibitem{Bergstrom:2014owa}
  J.~Bergstrom, D.~Meloni and L.~Merlo,
  Phys.\ Rev.\ D {\bf 89} (2014) no.9,  093021
  doi:10.1103/PhysRevD.89.093021
  [arXiv:1403.4528 [hep-ph]].

\bibitem{Hall:1999sn}
  L.~J.~Hall, H.~Murayama and N.~Weiner,
  Phys.\ Rev.\ Lett.\  {\bf 84} (2000) 2572
  doi:10.1103/PhysRevLett.84.2572
  [hep-ph/9911341].
  
  
\bibitem{Haba:2000be}
  N.~Haba and H.~Murayama,
  Phys.\ Rev.\ D {\bf 63} (2001) 053010
  doi:10.1103/PhysRevD.63.053010
  [hep-ph/0009174].
  
\bibitem{deGouvea:2003xe}
  A.~de Gouvea and H.~Murayama,
  Phys.\ Lett.\ B {\bf 573} (2003) 94
  doi:10.1016/j.physletb.2003.08.045
  [hep-ph/0301050].
  
\bibitem{Petcov:1982ya}
  S.~T.~Petcov,
  Phys.\ Lett.\  {\bf 110B} (1982) 245.
  doi:10.1016/0370-2693(82)91246-1
  
  
\bibitem{Altarelli:2005pj}
  G.~Altarelli and R.~Franceschini,
  JHEP {\bf 0603} (2006) 047
  doi:10.1088/1126-6708/2006/03/047
  [hep-ph/0512202].
  
\bibitem{Meloni:2011ac}
  D.~Meloni,
  JHEP {\bf 1202} (2012) 090
  doi:10.1007/JHEP02(2012)090
  [arXiv:1110.5210 [hep-ph]].
  
\bibitem{Lavoura:2000ci}
  L.~Lavoura and W.~Grimus,
  JHEP {\bf 0009} (2000) 007
  doi:10.1088/1126-6708/2000/09/007
  [hep-ph/0008020].
  
\bibitem{Grimus:2004cj}
  W.~Grimus and L.~Lavoura,
  J.\ Phys.\ G {\bf 31} (2005) 683
  doi:10.1088/0954-3899/31/7/013
  [hep-ph/0410279].
  
  
  
\bibitem{Altarelli:2015foa}
  G.~Altarelli, P.~A.~N.~Machado and D.~Meloni,
  PoS CORFU {\bf 2014} (2015) 012
  [arXiv:1504.05514 [hep-ph]].

  
  


\bibitem{Meloni:2011fx}
  D.~Meloni,
  JHEP {\bf 1110} (2011) 010
  doi:10.1007/JHEP10(2011)010
  [arXiv:1107.0221 [hep-ph]].
  

\bibitem{Altarelli:2008bg}
  G.~Altarelli, F.~Feruglio and C.~Hagedorn,
  JHEP {\bf 0803} (2008) 052
  doi:10.1088/1126-6708/2008/03/052
  [arXiv:0802.0090 [hep-ph]].

  
\bibitem{Kawamura:2000ev}
  Y.~Kawamura,
  Prog.\ Theor.\ Phys.\  {\bf 105} (2001) 999
  doi:10.1143/PTP.105.999
  [hep-ph/0012125].
  
\bibitem{Hall:2002ci}
  L.~J.~Hall and Y.~Nomura,
  Phys.\ Rev.\ D {\bf 66} (2002) 075004
  doi:10.1103/PhysRevD.66.075004
  [hep-ph/0205067].

\bibitem{Antusch:2010es}
  S.~Antusch, S.~F.~King and M.~Spinrath,
  Phys.\ Rev.\ D {\bf 83} (2011) 013005
  doi:10.1103/PhysRevD.83.013005
  [arXiv:1005.0708 [hep-ph]].
  
\bibitem{Antusch:2013wn}
  S.~Antusch, S.~F.~King and M.~Spinrath,
  Phys.\ Rev.\ D {\bf 87} (2013) no.9,  096018
  doi:10.1103/PhysRevD.87.096018
  [arXiv:1301.6764 [hep-ph]].
  
  
\bibitem{Chankowski:2001mx}
  P.~H.~Chankowski and S.~Pokorski,
  Int.\ J.\ Mod.\ Phys.\ A {\bf 17} (2002) 575
  doi:10.1142/S0217751X02006109
  [hep-ph/0110249].

\bibitem{Kuo:2001ha}
  T.~K.~Kuo, J.~T.~Pantaleone and G.~H.~Wu,
  Phys.\ Lett.\ B {\bf 518} (2001) 101
  doi:10.1016/S0370-2693(01)01032-2
  [hep-ph/0104131].
  
\bibitem{Antusch:2003kp}
  S.~Antusch, J.~Kersten, M.~Lindner and M.~Ratz,
  Nucl.\ Phys.\ B {\bf 674} (2003) 401
  doi:10.1016/j.nuclphysb.2003.09.050
  [hep-ph/0305273].
  
\bibitem{Antusch:2005gp}
  S.~Antusch, J.~Kersten, M.~Lindner, M.~Ratz and M.~A.~Schmidt,
  JHEP {\bf 0503} (2005) 024
  doi:10.1088/1126-6708/2005/03/024
  [hep-ph/0501272].
  

  
  
\bibitem{Hagedorn:2012ut}
  C.~Hagedorn, S.~F.~King and C.~Luhn,
  Phys.\ Lett.\ B {\bf 717} (2012) 207
  doi:10.1016/j.physletb.2012.09.026
  [arXiv:1205.3114 [hep-ph]].
  
\bibitem{Cooper:2012wf}
  I.~K.~Cooper, S.~F.~King and C.~Luhn,
  JHEP {\bf 1206} (2012) 130
  doi:10.1007/JHEP06(2012)130
  [arXiv:1203.1324 [hep-ph]].

\bibitem{Marzocca:2011dh}
  D.~Marzocca, S.~T.~Petcov, A.~Romanino and M.~Spinrath,
  JHEP {\bf 1111} (2011) 009
  doi:10.1007/JHEP11(2011)009
  [arXiv:1108.0614 [hep-ph]].

\bibitem{Antusch:2013kna}
  S.~Antusch, C.~Gross, V.~Maurer and C.~Sluka,
  Nucl.\ Phys.\ B {\bf 877} (2013) 772
  doi:10.1016/j.nuclphysb.2013.11.003
  [arXiv:1305.6612 [hep-ph]].
  
\bibitem{Antusch:2013tta}
  S.~Antusch, C.~Gross, V.~Maurer and C.~Sluka,
  Nucl.\ Phys.\ B {\bf 879} (2014) 19
  doi:10.1016/j.nuclphysb.2013.11.017
  [arXiv:1306.3984 [hep-ph]].

  
\bibitem{Bjorkeroth:2015ora}
  F.~Björkeroth, F.~J.~de Anda, I.~de Medeiros Varzielas and S.~F.~King,
  JHEP {\bf 1506} (2015) 141
  doi:10.1007/JHEP06(2015)141
  [arXiv:1503.03306 [hep-ph]].

\bibitem{Antusch:2017ano}
  S.~Antusch and C.~Hohl,
  arXiv:1706.04274 [hep-ph].

  
\bibitem{Gehrlein:2014wda}
  J.~Gehrlein, J.~P.~Oppermann, D.~Schäfer and M.~Spinrath,
  Nucl.\ Phys.\ B {\bf 890} (2014) 539
  doi:10.1016/j.nuclphysb.2014.11.023
  [arXiv:1410.2057 [hep-ph]].

  
\bibitem{Meroni:2012ty}
  A.~Meroni, S.~T.~Petcov and M.~Spinrath,
  Phys.\ Rev.\ D {\bf 86} (2012) 113003
  doi:10.1103/PhysRevD.86.113003
  [arXiv:1205.5241 [hep-ph]].
  


\bibitem{Altarelli:2010at}
  G.~Altarelli and G.~Blankenburg,
  JHEP {\bf 1103} (2011) 133
  doi:10.1007/JHEP03(2011)133
  [arXiv:1012.2697 [hep-ph]].

\bibitem{Joshipura:2009tg}
A.~S.~Joshipura, B.~P.~Kodrani and K.~M.~Patel,
Phys.\ Rev.\ D {\bf 79} (2009) 115017
doi:10.1103/PhysRevD.79.115017
[arXiv:0903.2161 [hep-ph]].

\bibitem{Grimus:2006rk}
W.~Grimus and H.~Kuhbock,
Eur.\ Phys.\ J.\ C {\bf 51} (2007) 721
doi:10.1140/epjc/s10052-007-0324-5
[hep-ph/0612132].
  
\bibitem{Dermisek:2005ij}
R.~Dermisek and S.~Raby,
Phys.\ Lett.\ B {\bf 622} (2005) 327
doi:10.1016/j.physletb.2005.07.018
[hep-ph/0507045].

\bibitem{Dermisek:2006dc}
R.~Dermisek, M.~Harada and S.~Raby,
Phys.\ Rev.\ D {\bf 74} (2006) 035011
doi:10.1103/PhysRevD.74.035011
[hep-ph/0606055].
  
\bibitem{Albright:2000dk}
C.~H.~Albright and S.~M.~Barr,
Phys.\ Rev.\ D {\bf 62} (2000) 093008
doi:10.1103/PhysRevD.62.093008
[hep-ph/0003251].

\bibitem{Albright:2001uh}
C.~H.~Albright and S.~M.~Barr,
Phys.\ Rev.\ D {\bf 64} (2001) 073010
doi:10.1103/PhysRevD.64.073010
[hep-ph/0104294].
  
\bibitem{Ji:2005zk}
X.~d.~Ji, Y.~c.~Li and R.~N.~Mohapatra,
Phys.\ Lett.\ B {\bf 633} (2006) 755
doi:10.1016/j.physletb.2006.01.005
[hep-ph/0510353].


\bibitem{Bjorkeroth:2017ybg}
  F.~Björkeroth, F.~J.~de Anda, S.~F.~King and E.~Perdomo,
  arXiv:1705.01555 [hep-ph].


\bibitem{Bjorkeroth:2015uou}
  F.~Björkeroth, F.~J.~de Anda, I.~de Medeiros Varzielas and S.~F.~King,
  Phys.\ Rev.\ D {\bf 94} (2016) no.1,  016006
  doi:10.1103/PhysRevD.94.016006
  [arXiv:1512.00850 [hep-ph]].
  
  
  
  
\bibitem{Bjorkeroth:2016lzs}
  F.~Björkeroth, F.~J.~de Anda, I.~de Medeiros Varzielas and S.~F.~King,
  JHEP {\bf 1701} (2017) 077
  doi:10.1007/JHEP01(2017)077
  [arXiv:1609.05837 [hep-ph]].


\bibitem{Toorop:2010yh}
  R.~de Adelhart Toorop, F.~Bazzocchi and L.~Merlo,
  JHEP {\bf 1008} (2010) 001
  doi:10.1007/JHEP08(2010)001
  [arXiv:1003.4502 [hep-ph]].

\bibitem{King:2013hoa}
  S.~F.~King,
  JHEP {\bf 1401} (2014) 119
  doi:10.1007/JHEP01(2014)119
  [arXiv:1311.3295 [hep-ph]].
  
\bibitem{King:2014iia}
  S.~F.~King,
  JHEP {\bf 1408} (2014) 130
  doi:10.1007/JHEP08(2014)130
  [arXiv:1406.7005 [hep-ph]].

\bibitem{King:2006me}
  S.~F.~King and M.~Malinsky,
  JHEP {\bf 0611} (2006) 071
  doi:10.1088/1126-6708/2006/11/071
  [hep-ph/0608021].

  
\bibitem{Adamson:2017gxd}
  P.~Adamson {\it et al.} [NOvA Collaboration],
  Phys.\ Rev.\ Lett.\  {\bf 118} (2017) no.23,  231801
  doi:10.1103/PhysRevLett.118.231801
  [arXiv:1703.03328 [hep-ex]].

\bibitem{Abe:2017uxa}
  K.~Abe {\it et al.} [T2K Collaboration],
  Phys.\ Rev.\ Lett.\  {\bf 118} (2017) no.15,  151801
  doi:10.1103/PhysRevLett.118.151801
  [arXiv:1701.00432 [hep-ex]].

\bibitem{Acciarri:2016crz}
  R.~Acciarri {\it et al.} [DUNE Collaboration],
  arXiv:1601.05471 [physics.ins-det].

\bibitem{Djurcic:2015vqa}
  Z.~Djurcic {\it et al.} [JUNO Collaboration],
  arXiv:1508.07166 [physics.ins-det].

\end{thebibliography}

\end{document}